\documentclass[manuscript,screen,review]{acmart}
\AtBeginDocument{%
  \providecommand\BibTeX{{%
    \normalfont B\kern-0.5em{\scshape i\kern-0.25em b}\kern-0.8em\TeX}}}

\setcopyright{acmcopyright}
\copyrightyear{2022}
\acmYear{2022}
\acmDOI{10.1145/1122445.1122456}

\usepackage{paralist}

\begin{document}

\title{Survey on Privacy-Preserving Techniques for Data Publishing}

\author{Tânia Carvalho}
\email{tania.carvalho@fc.up.pt}
\orcid{0000-0002-7700-1955}
\affiliation{%
  \institution{DCC - Faculty of Sciences, University of Porto}
  \city{Porto}
  \country{Portugal}
}

\author{Nuno Moniz}
\affiliation{%
  \institution{INESC TEC / University of Porto}
  \city{Porto}
  \country{Portugal}
}
\email{nmmoniz@inesctec.pt}
\orcid{0000-0003-4322-1076}

\author{Pedro Faria}
\affiliation{%
  \institution{TekPrivacy}
  \city{Porto}
  \country{Portugal}
}
\email{pvfaria@tekprivacy.com}

\author{Luís Antunes}
\affiliation{%
  \institution{DCC - Faculty of Sciences, University of Porto}
  \city{Porto}
  \country{Portugal}
 }
\email{lfa@fc.up.pt}
\orcid{0000-0002-9988-594X}

\renewcommand{\shortauthors}{Carvalho et al.}

\begin{abstract}
The exponential growth of collected, processed, and shared microdata has given rise to concerns about individuals' privacy. As a result, laws and regulations have emerged to control what organisations do with microdata and how they protect it. Statistical Disclosure Control seeks to reduce the risk of confidential information disclosure by de-identifying them. Such de-identification is guaranteed through privacy-preserving techniques. However, de-identified data usually results in loss of information, with a possible impact on data analysis precision and model predictive performance. The main goal is to protect the individuals' privacy while maintaining the interpretability of the data, i.e. its usefulness. Statistical Disclosure Control is an area that is expanding and needs to be explored since there is still no solution that guarantees optimal privacy and utility. This survey focuses on all steps of the de-identification process. We present existing privacy-preserving techniques used in microdata de-identification, privacy measures suitable for several disclosure types and, information loss and predictive performance measures. In this survey, we discuss the main challenges raised by privacy constraints, describe the main approaches to handle these obstacles, review taxonomies of privacy-preserving techniques, provide a theoretical analysis of existing comparative studies, and raise multiple open issues.
\end{abstract}



\maketitle

\section{Introduction}

The right to privacy is a discussion topic since the late 60s in the computing field~\cite{hoffman1969computers}. At that time, awareness of individuals' privacy marked the development of legal and administrative safeguards appropriate to the computerised and modern world. Privacy is therefore not a recent topic of concern and debate. However, today, we face the pressing problem of data pervasiveness and the issues it raises for computation and society. The amount of collected and shared data reached astonishing levels, providing detailed records of information on individuals and used in real-world applications. This type of data -- microdata -- is often seen as crucial for analysis and mining tasks in data-driven projects. Microdata is also key to research and strategy development efforts, as to gather meaningful information and in-depth knowledge from available sources. Nonetheless, despite the critical importance and vast range of advantages in information sharing, the privacy of data subjects is constantly challenged as data is re-used and analysed on an unprecedented scale. As a result, privacy raised multiple legal puzzles around the world, heightening the interest and concern of individuals about what organisations and institutions handle their data and private information.

Recently, several data privacy regulations have been put in place to protect data subjects' privacy. The General Data Protection Regulation\footnote{Regulation (EU) 2016/679 of the European Parliament and of the Council of 27 April 2016} (GDPR) emerged to unify data privacy laws across Europe. Following the GDPR, several efforts were put in place~\cite{rustad2019towards}, such as the California Consumer Privacy Act of 2018 (CCPA) and the General Personal Data Protection Law (Lei Geral de Proteção de Dados Pessoais, or LGPD) were passed to protect the privacy of Californian and Brazilian individuals, respectively. In essence, these pieces of legislation require entities to establish appropriate technical and organisational measures to process personal data in compliance with such laws and regulations. Legally, a data subject is an individual to whom data relates. Data controllers correspond to entities which determine the purposes and means of the processing of personal data. Data processors handle personal data on behalf of controllers~\cite{edpb_datacontroller_processor}. While the data subject has rights over its data, namely rights of access or erasure, the data controller is subject to various obligations, such as ensuring confidentiality, notifying if data are breached and carrying out risk assessments~\cite{wp29_breach}. 

To ensure data confidentiality is primarily to prevent private information disclosure, by limiting data access to authorised entities or by de-identifying the data, i.e., all private information concerning an individual in a record or data set is removed or transformed. The de-identification procedure reduces the amount of information and data granularity, which typically results in losses for predictive and/or descriptive performance, as well as in data interpretability~\cite{brickell2008cost, li2009tradeoff}. In the case of machine learning/data mining tasks, we face what can be denoted as a trade-off between privacy and predictive performance~\cite{carvalho2022data}. It is fundamental to design privacy-preserving techniques that best guarantee privacy without compromising (or compromising as little as possible) predictive performance. However, finding the best trade-off remains a challenge. On one hand, poor-quality data (high privacy level) makes the interpretation of results and extraction of knowledge less effective. On the other hand, low privacy level may result in re-identification. Without addressing these obstacles, the outcome of both approaches may be skewed. For example, statistical analysis may produce wrong conclusions; and, data breaches may result in identity theft used for fraudulent purposes. 

There are two major research topics focusing on the protection of individuals' privacy which are of interest to our work: data encryption and data transformation. The first, data encryption, is mainly related to the use of cryptographic methods. However, a major drawback of such type of methods is the subsequent difficulty in data manipulation, analysis and results' interpretation. The second, data transformation, encapsulates all methods intended to transform and prepare data for public availability, research or industrial usage. These two groups of methods are commonly associated to distinct underlying conditions. For example, data encryption is useful when third parties cannot be trusted, the organisation does not need the real data for business operations or the compliance standards require encryption under specific conditions. As for data transformation, such methods are of crucial importance when interpretability is a critical factor, and the objective is to share meaningful information while preserving the privacy of individuals. In Statistical Disclosure Control, data transformation is key to ensure the protection of the data when released, i.e. de-identified, where such transformation is achieved through privacy-preserving techniques.


In this survey, we review the Statistical Disclosure Control problem and present a survey on privacy-preserving techniques for data transformation. We focus on methods that aim at preserving data and results' interpretability, as well as their ability to serve as the basis for future analysis, motivated by the importance of secure data sharing for future endeavours in computation. The main contributions of this work are summarised as follows: \textit{i)} provide a general definition of the de-identification problem in microdata; \textit{ii)} describe the main privacy risk measures concerning specific disclosure type of private information; \textit{iii)} propose a taxonomy of existing approaches of privacy-preserving techniques in data transformation; \textit{iv)} describe the well-known measures of information loss and predictive performance; \textit{v)} summarise the conclusions of existing experimental comparisons, and; \textit{vi)} review theoretical problems of de-identification processes. Concerning previous work, we should stress that, despite the importance of privacy-preserving techniques in microdata, there is no recent overview and discussion on this topic. In addition, existing surveys only address a part of privacy-preserving techniques and privacy risk measures for a certain type of disclosure. Furthermore, such studies do not address the effectiveness of privacy-preserving techniques (e.g.~\cite{domingo2008survey, fung2010privacy}). Our survey builds on previous work, by discussing and updating existing taxonomies for privacy-preserving techniques, including more recent techniques. Then, we provide a thorough discussion on privacy risks and review important advancements. Finally, we provide an extensive analysis of the effectiveness of such techniques concerning privacy protection and predictive performance along with a summary of the main conclusions of existing experimental studies.

The remainder of the article is organised as follows. Section~\ref{sec:prob} includes some preliminaries and defines the problem of Statistical Disclosure Control. Section~\ref{sec:SDCprocess} describes each stage in de-identification process. Section~\ref{sec:privlevel} details several measures to assess the privacy risk according to different disclosure types. Section~\ref{sec:PPT} provides a taxonomy for the set of privacy-preserving techniques as well as their description. Section~\ref{sec:utility} presents an overview of the main approaches for utility assessment. In Section~\ref{sec:studies} we present some existing experimental comparisons of different strategies for this problem. Section~\ref{sec:openissues} explores some problems related to Statistical Disclosure Control and includes a summary of recent trends and open research questions. Finally, Section~\ref{sec:conslusion} concludes the article.

\section{Preliminaries and Problem Formulation}\label{sec:prob}

The Statistical Disclosure Control (SDC) task, many times referred to as Statistical Disclosure Limitation or Inference Control, is a principle that aims to provide statistical data to society while preserving data subjects' privacy. De-identified statistical data is achieved by using privacy-preserving techniques in such a way that is impossible to disclose any confidential information on any data subject. To determine which privacy-preserving techniques are appropriate for data protection and potential threats, it is essential to distinguish between database formats, because each presents different challenges. In this context, microdata, tabular data and query-based databases are the most common formats, as outlined by~\citet{enisa}. Microdata consists of a set of records where each entry contains information that corresponds to a specific individual. Tabular data represents aggregated information for specific individual groups, which may include counts or magnitudes. 
Lastly, query-based databases are iterative databases where users submit statistical queries such as sums, averages, max, min, and others. 
    
    
    
    
    

One should note that microdata sets are the raw material used to construct both tabular data and query-based databases. Accordingly, henceforth, we focus on the principles applied to protect microdata and the main privacy measures used for privacy guarantees. Furthermore, the disclosure risk for microdata is potentially high when compared with tables~\cite{willenborg1996statistical}, which is an important reason for increasing attention to this type of data. Nevertheless, we should note that the principles to protect tabular data has been surveyed by many researchers~\cite{willenborg1996statistical, willenborg2000elements, hundepool2012statistical, domingo2016database}. Also, concerning query-based databases,~\citet{adam1989security} present a comparative study to protect this type of data. 
Focusing on the microdata setting, attributes traditionally obey the following terminology.

\begin{itemize}
    \item[--] \textit{Identifiers}: attributes such as name and social security number that directly identify an individual.

    \item[--] \textit{Quasi-identifiers (QI)}: attributes that, when combined, generate a unique signature that may lead to re-identification. For instance, date of birth, gender, geographical location, profession and ethnic group. In related literature, these are frequently called key attributes.

    \item[--] \textit{Sensitive}: known as confidential attributes, they refer to highly critical attributes, usually protected by law and regulations. For example, religion, sexual orientation, disease and political opinion.

    \item[--] \textit{Non-sensitive}: other attributes that do not contain sensitive information. 

\end{itemize}

The problem addressed in SDC is that, through inappropriate use, a person who is given access to released data may disclose private information about data subjects. Due to such liability, identity disclosure is one of the main concerns of data privacy today. An intruder, also known as an adversary, attacker or snooper, is an individual who possesses the skills, resources and motivation to re-identify data subjects or deduce new information about them in the de-identified data set. Motivated intruders can improve their knowledge of private information on observations in available data. It is fundamental for data controllers to make assumptions regarding intruders' background knowledge. If an intruder has more background knowledge than assumed by data controllers, the risk of disclosure may be underestimated.

Several studies show how feasible and/or easy it is to link back private information to a data subject~\cite{re-id-homicides, Netflix, rocher2019estimating, carvalho2021fundamental}. Despite the possibility of linking information derived from different sources, the re-identification can also occur by isolating data subjects in the de-identified data. In a study conducted by~\citet{sweeney2000simple}, it was found that 87\% of the population in the United States are likely to be uniquely identified by only considering the set of QI \{5-digit ZIP, gender, date of birth\}. Such a study contributed to the increased attention from both data controllers and data subjects and was an inspiration to create new techniques for data protection and measures to minimise disclosure risk. 

\begin{figure}[!ht]
   \centering
   \scriptsize
   \includegraphics[width=0.4\linewidth]{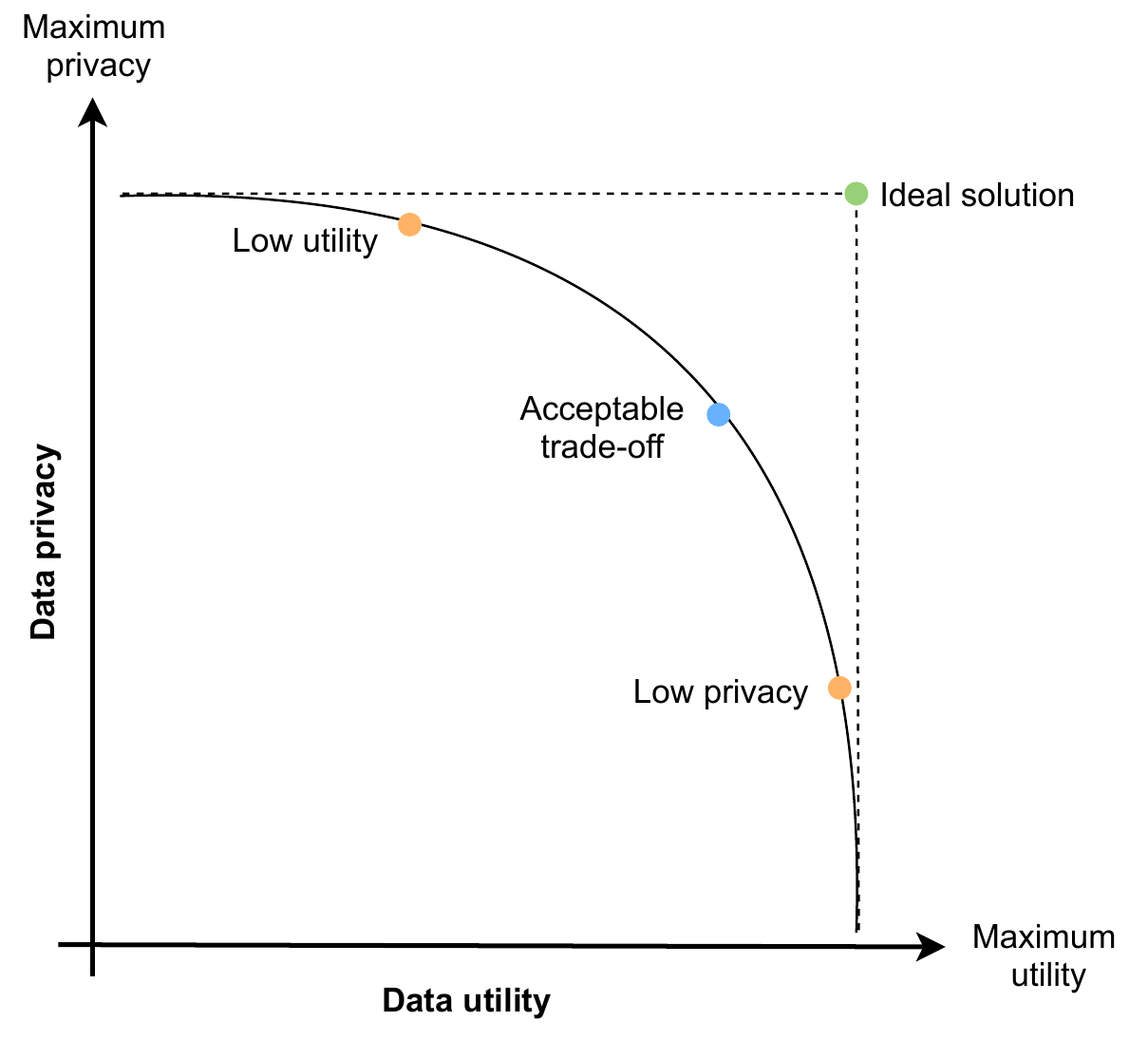}
 \caption{Trade-off between privacy level and utility level of data.}
 \label{fig:tradeoff}
\end{figure}

The design of robust privacy-preserving techniques includes assessing the impact of such techniques for privacy and interpretability/utility. Notably, they must guarantee desired levels of data protection without compromising the usefulness of data. Figure~\ref{fig:tradeoff} illustrates an acceptable trade-off between data privacy and utility. Although the ideal solution is de-identified data with maximum privacy and utility, this scenario is practically impossible to reach. 

The formal definition of the presented problem is described as follows. Consider a microdata set $T = \{t_1, ..., t_n\}$, where $t_i$ corresponds to a tuple of attribute values for an individuals' record. Let $V = \{v_1, ..., v_m\}$ be the set of $m$ attributes, in which $t_{i,v_j}$ denotes the value of attribute $v_j$ for tuple $i$. A QI consists of a set of attribute values (either categorical or numeric) that could be known to the intruder for a given individual where $QI \in V$. An equivalence class corresponds to a combination of records that are indistinguishable from each other. Then, two tuples, $t_a$ and $t_b$ are in same equivalence class if $t_{a,[QI]} = t_{b,[QI]}$. In the de-identification process, one or more ``transformation functions'' are applied to the original data, producing de-identified data. A data set with minimum privacy guarantees does not have equivalence classes of size one, i.e. a distinct set of values in a group of attributes for a single individual, and an intruder cannot isolate any data points or infer any information about an individual. 

Although privacy-preserving techniques aim to transform confidential information, it may not be enough to protect it from intruders linking private information back to an individual or singling out individuals that have indistinguishable information. In addition to the presentation of the main privacy-preserving techniques and measures for disclosure risk and data utility, we review and discuss available studies concerning the effectiveness of such techniques on both privacy and, when available, predictive performance in data mining/machine learning tasks.

\section{De-identification Process}\label{sec:SDCprocess}

As stated before, SDC in a microdata setting ensures the release of a de-identified data set without compromising confidential information that can be linked to specific data subjects or entities. Such de-identified data is achieved through a de-identification process which may produce multiple statistical outputs.
Different outputs require different approaches to SDC as it depends on a specific end-use~\cite{SDCforBusiness}. The common purposes for microdata are outlined as follows. 

\begin{itemize}
    \item[--] \textit{Secure use files}: the data is available in a safe centre, usually controlled by national statistical authorities. In this type of files, the level of disclosure risk is high because only direct identifiers are treated.
    
    \item[--] \textit{Scientific use files}: the data can be used outside of controlled environments. The risk level is medium and the security of the data is the responsibility of the data receiver that is usually a researcher. 
    
    \item[--] \textit{Industry use files}: the difference from scientific files is the destination of this type of data. The data is released for industrial usage, related for instance with telecommunication or insurance domains.
    
    \item[--] \textit{Public use files}: the data is available to the general public. The level of protection in this type of files are very high to prevent the disclosure of private information. The risk is very low (or null) under specific attacker scenarios.   
    
\end{itemize}


Secure, scientific and public use files are the three best-known microdata destinations~\cite{SDCforBusiness}. We introduce a new end-use for microdata: industry use files. This type of file concerns scenarios where an organisation needs to extract knowledge and add value to their business. In such scenarios, data can be used outside a controlled environment as with scientific public use files. Disclosure risk is also the responsibility of data receivers.

\begin{figure}[ht!]
  \centering
  \scriptsize
  \includegraphics[width=0.5\linewidth]{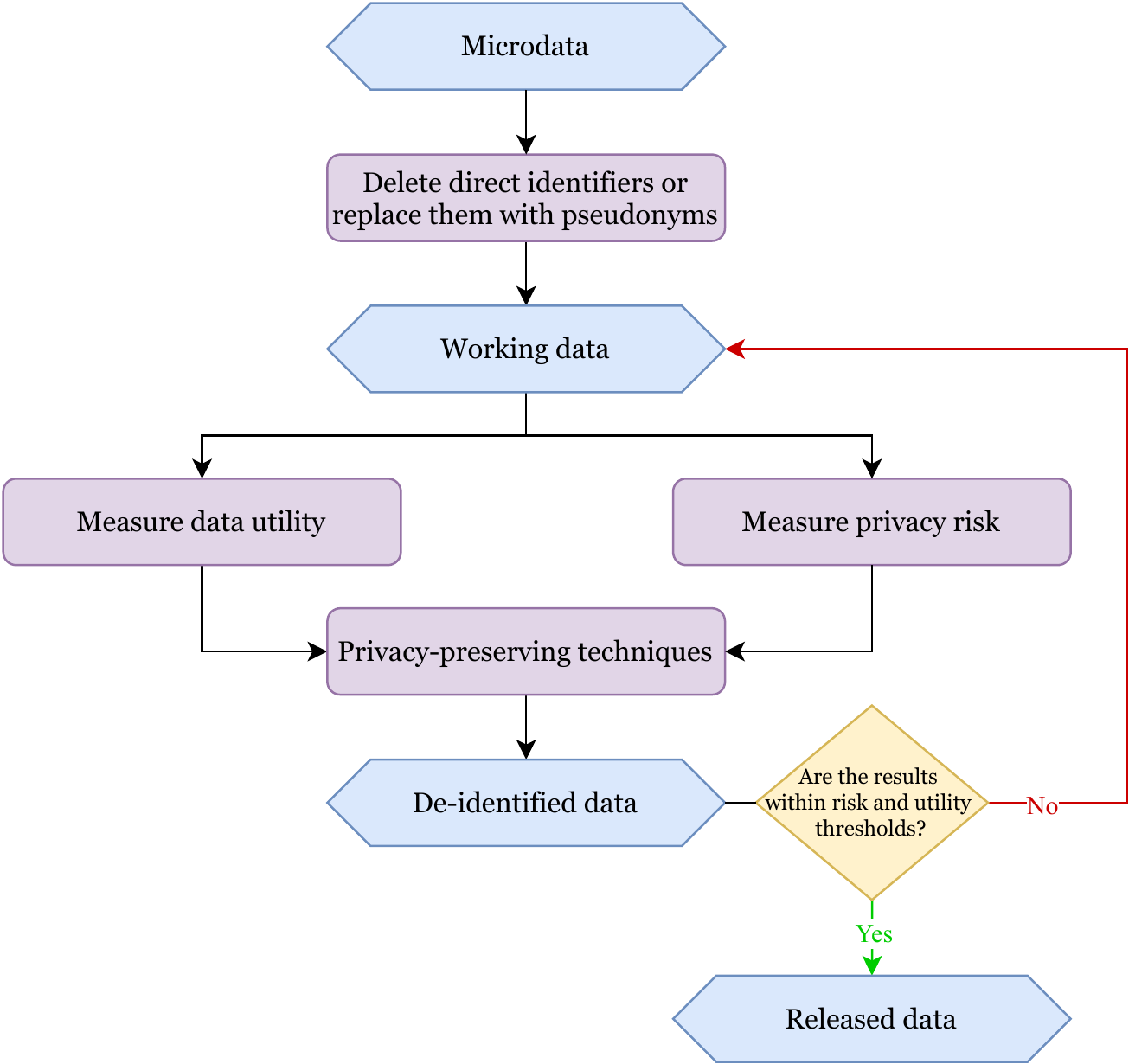}
\caption{De-identification process for microdata.}
\label{fig:flux}
\end{figure}

The destination of microdata is then a crucial factor for determining de-identification levels as well as the requirements end-users have for the data. For scientific, industry and public purposes, microdata must be de-identified by applying privacy-preserving techniques, while ensuring a balance between data privacy and interpretability/utility. A general overview of the main steps in the de-identification process applied for microdata is presented in Figure~\ref{fig:flux}. 

The first step is to identify the direct identifiers since such attributes are extremely risky as they easily expose the identity of a data subject. Then, direct identifiers must be removed or replaced with pseudonyms. The remaining attributes are used to assess the raw disclosure risk and utility. The choice of privacy-preserving techniques is based on the need for data protection determined by the disclosure risk and structure of data. After data de-identification, the disclosure risk and utility is re-assessed. If the compromise between the two measures is not met, the parameters of privacy-preserving techniques should be re-adjusted or different techniques must be applied. The process is repeated until the desired level of privacy and utility is reached. Otherwise, the data is protected and can be released. 


It is fundamental to have documentation on the de-identification process, namely for auditing from specialised authorities. The documentation should include a description of legal and administrative safeguards for risk management and technical solutions employed. Furthermore, such documentation is crucial for transparency towards data subjects and data receivers to understand what was changed in the original data due to confidentiality constraints. It may be useful for data subjects or data receivers to know which attributes were modified along with a brief description of changes either by privacy-preserving techniques application or out of necessity due to the quality of original data. Notwithstanding, in data mining, a recent suggestion was proposed for documentation to transparent model reporting~\cite{mitchell2019model}. The goal is to clarify the potential use cases of trained machine learning models and minimise their usage in contexts for which they are not appropriate. For model transparency, an important criteria to be considered is the ``ethical consideration'' that includes the evidence of risk mitigation and the risk and harms present in the model.

De-identification effectiveness is determined by how likely an intruder will be able to disclose personal information on any data subject. Therefore, in the following sections, we thoroughly describe disclosure risk, privacy-preserving techniques used to circumvent problems caused by disclosure of confidential information, and finally, existing measures to boost data interpretability and utility for data mining/machine learning tasks.  

\section{Privacy Risk} \label{sec:privlevel}

Considering the terminology for attributes in microdata, robust privacy-preserving techniques must be selected by data controllers to obtain de-identified data sets. Such transformation is commonly based on three common types of privacy threats~\cite{wp29}: \textit{i)} \textbf{singling out}, where an intruder can isolate records that uniquely identify a data subject; \textit{ii)} \textbf{linkability}, concerning the ability to connect or correlate two or more records of an individual or a group of individuals, and \textit{iii)} \textbf{inference}, regarding the possibility to deduce the value of an attribute based on the values of other attributes. We should also note, increased attempts on confidential information attacks have contributed to the awareness of other privacy threats, which we also discuss, despite them being less known. 

A data controller usually validates the effectiveness of a privacy-preserving technique through privacy measures appropriated for each previous privacy threat. Quantifying disclosure risk is a difficult task because disclosure of confidential information generally occurs when an intruder possesses external information, and the data controller often cannot anticipate this information. Therefore, a data controller needs to prudently make assumptions about her knowledge to predict disclosure risk. Usually, the data controller examines risks under diverse scenarios, for instance, different sets of QI known by intruders, or whether or not intruders know who participated in the study. Unfortunately, assumptions may not be accurate for a given de-identified data set. To prevent this, the best approach is to assume a maximum-knowledge intruder, i.e. the intruder knows all original attribute values. Hence, she assumes that an intruder may use all QI (maximum background knowledge), which allows her to obtain the most accurate risk estimate possible.

In the following sections, we specify four types of disclosure risks described in the literature and their respective measures: identity, attribute, inferential and membership. 

\begin{itemize}
     \item[--] \textbf{Identity disclosure:} when an intruder can recognise that a record in the released data set concerns an individual by matching QI values;
    \item[--] \textbf{Attribute disclosure:} when an intruder is able to determine new characteristics of a data subject based on the information available in the released data;
    \item[--] \textbf{Inferential disclosure:} when an intruder can infer data subject private information with high confidence from statistical properties of the released data;
    \item[--] \textbf{Membership disclosure:} when an intruder is able to conclude if the private information regarding a certain individual is present or not in the data set.  
    
\end{itemize}

We should stress that inferential disclosure risk requires a data mining technique performed by a data processor to unlawfully gain knowledge about a data subject. In other words, an intruder can predict the value of a certain individual's characteristics more accurately with released data~\cite{hundepool2012statistical}. Inferential disclosure may harm a group of individuals, even individuals whose information do not appear in the data set~\cite{domingo2016database}. As inference strategies are designed to predict aggregate, not individual behaviour or attributes, this is not addressed in SDC for the microdata setting. Accordingly, we do not discuss measures for inferential disclosure in this survey. Also, to our current knowledge, there are only a few measures for membership disclosure risk, such as $\delta$-presence~\cite{nergiz2007hiding}, aiming to bound the likelihood of inferring the presence of any potential data subjects' record within a specified range $\delta=(\delta _{min}, \delta _{max})$. Due to the imbalanced representation of this type of disclosure risk, we will also not delve into its respective measures in the following sections. 
 

\subsection{Identity Disclosure}

There are two main strategies for measuring identity disclosure risks in microdata~\cite{reiter2005estimating}: \textit{i)} estimating the number of records released in the sample whose characteristics are unique in the population (uniqueness), and \textit{ii)} estimating the probability that records possessed by intruders can be identified from the released data (record linkage). Notwithstanding, some strategies used for other purposes have been adapted to find records at risk, and for this reason, we also include measures for \textit{iii)} outlier detection and \textit{iv)} clustering.

\subsubsection{Uniqueness}

Population uniqueness is essential for successful disclosure. Suppose that a data receiver knows that a specific individual is unique in the population. Then, this individual either is or not in the sample. If the former, such an individual will be identified and disclosed with high certainty of de-identification; if the latter, the individual is not in the sample and no harm can be done. For this reason, uniqueness is very relevant, since \textit{population uniques} naturally have a higher risk of re-identification than non-uniques. Thus, knowledge of population uniqueness should not be underestimated, especially when microdata sets contain attributes that make it possible to easily detect individuals. For instance, certain professions could be unique in a small geographical areas. Furthermore, in case the sample size is equivalent to the population size, an intruder that detects a unique value in the released sample is almost certain that there is a single individual in the population with that value, making it identifiable.

We assume three different procedures to determine uniqueness. First, by isolating records according to a set of QI. Second, the data receiver may not have the resources to acquire data on all of the population and she needs to estimate uniqueness from the available sample data by using a probabilistic model. Lastly, an alternative is based on special uniques' detection which searches for uniqueness without considering the entire QI set. All procedures are important to decide whether re-identification risk is acceptable or if further disclosure control actions are required. Therefore, we discuss the following group of measures for uniqueness: singling out, probabilistic modelling, and special uniques.

\textbf{Singling Out.} When all QI of a de-identified data set are categorical, estimation of disclosure risk can be obtained by the frequency of a set of QI. Let $C_1, ..., C_K$ represent $K$ records, and the population and sample frequencies $F_k=(k=1, ..., F_K)$ and $f_k=(k=1, ..., f_K)$, respectively. Formally, $\sum_{k=1}^K F_k=N$ and $\sum_{k=1}^K f_k=n$ where $N$ and $n$ correspond to the size of population and sample, respectively. The probability of identity disclosure of an individual $i$ being in cell $C_k$, when $F_k$ individuals in the population are known to belong to it, is $1/F_k$ for $k=1, ..., K$. If $F_k=1$ then the combination of QI values is unique in the population. As such, the intruder is sure that matching record for individual $i$ on the QI does indeed belong to that individual. This means that the intruder can single out the correspondent individual. If $F_k=2$ then the re-identification risk is $0.5$ which means that the intruder has 50\% of certainty on the re-identification.

A very common criterion for deciding whether the risk of re-identification is too high is $k$-anonymity, proposed by~\citet{samarati2001protecting}. Such a measure and the many extensions based on $k$-anonymity are referred to in the literature as ``privacy models''. For instance, in the case of $k$-anonymity, researchers use those ``models'' to indicate whether a data set respects the desired level of $k$ to protect against background knowledge. Along with that, this type of approach provides the level of disclosure risk. Therefore, for terminology clarity, we assume that $k$-anonymity and its variations are measures and not models, i.e. they are mainly used in performance assessment and/or criteria in optimisation processes. Many surveys detail several measures related to $k$-anonymity and provide an analysis of their strengths and weaknesses~\cite{fung2010privacy, fung2010introduction, zigomitros2020survey, majeed2020anonymization}. Regarding $k$-anonymity, it is similar to the previous sample uniqueness definition. Each equivalence class is assigned a frequency $f_k$. A record is unique when $k=1$. For better protection, $k$ must be great than 1. It is not always appropriate to use $k$-anonymity, for example when this metric indicates that a certain record is unique but there is not enough information to re-identify it. Therefore, is needed more information: a re-identification dataset. $k$-map~\cite{el2008protecting} arose to address this shortcoming of $k$-anonymity. Both measures are very similar with the exception that $k$-map calculates the risk based on information about the underlying population.


Finally, determining population uniqueness requires access to population data which is rarely available. Nonetheless, when population frequencies $F_k$ are unknown, they can be estimated from the sample with statistical models~\cite{hundepool2010handbook}. For example, the objective of super-population models is to estimate the characteristics of the overall population using probability distributions that are parameterised with sample characteristics.

\textbf{Probabilistic Modelling.} ~\citet{bethlehem1990disclosure} proposes a model for estimating the number of population uniques using sample data based on the assumption that cell frequencies are a realisation of a super-population (a theoretical population) distribution. In a population with $N$ individuals and $K$ cells, each cell $k$ is assigned a super-population parameter $\pi>0$ (a probability) and an attribute $F_k$ with the population frequency in that cell. It is assumed that $F_k$ follows a Poisson distribution with expected value $\mu_k=N\pi_k$. Thus, the expected number of population uniques ($U_p$) is denoted as $U_p=\sum_{k=1}^K\mu_k exp(-\mu_k)$ which can be used as an approximation to the realised number of unique individuals under the super-population model. However, estimating all expected values is a complex problem due to the large number of cells. The Poisson-gamma model serves to govern the generation of the super-population parameters by considering $\pi_k$ as a realisation of gamma($\alpha, \beta$)-distributed attributes denoted by $\Pi_k$, where $\alpha=1/K\beta$ and $\beta$ reflects the amount of dispersion of $\Pi_k$. Thus, the Poisson-gamma model is summarised as $F_k \sim Poisson(N\pi_k)|\pi_k=\Pi_k$ and $\Pi_k \sim gamma(\alpha,\beta)$. To estimate $U_p$, we require the parameters $\alpha$ and $\beta$, which can be given by maximum likelihood estimators. Also, a well-known model in the literature is the Poisson-log-normal which considers how an intruder might use released microdata sets to infer whether a sample unique record is population unique~\cite{skinner1998estimating, skinner2002measure}. The measure depends on the specification of a log-linear generalisation of a Poisson log-normal model for a set of QI. 

\citet{hoshino2001applying} proposes the use of Pitman's model for the same purpose. Such a model is defined in terms of the cell size indices and is a generalisation of Ewens sampling formula~\cite{ewens1990population}. Additional to Poisson-gamma and Poisson-log-normal models, the authors present a comparison including the Dirichlet-multinomial model~\cite{takemura1999some}, logarithmic series model~\cite{hoshino2001applying} and Ewens model. According to their results, the most accurate result was obtained with Pitman's model.~\citet{dankar2012estimating} have experimentally validated Pitman's sampling formula as the underlying distribution with clinical datasets. Furthermore, the Pitman's model has been employed in the well-known privacy tool ARX~\cite{prasser2016importance}. 

A different model to compute the population uniqueness is presented by~\citet{rocher2019estimating}. Gaussian copulas are used to model population uniqueness and estimate the likelihood for a sample unique being a population unique. The model quantifies, for any individual $i$, the likelihood $\xi _i$ for this record to be unique in the complete population. From $\xi _i$, it is derived the likelihood $\upsilon _i$ for $i$ to be correctly re-identified when matched. The Gaussian copulas allow modelling the density of probability distributions by specifying separately the marginal distributions and the dependency structure. 

\textbf{Special Uniques.} A different approach is based on the concept of a special unique~\cite{elliot2002computational}. A special unique has a higher probability of being a population unique than a normal sample unique. Within sample uniques w.r.t a set of QI, it is possible to find unique patterns without even considering the complete set of QI. The subset of QI is referred to as the Minimal Sample Unique (MSU) as any smaller subset of this set is not unique. The method was implemented into Special Uniques Detection Algorithm (SUDA). To fulfil the minimal requirement, all subsets of size $k-1$ of the MSU are unique. The principal objective of SUDA is then to identify all the MSUs in the sample. The potential risk of the records is determined based on the size of MSU; the smaller the size, the greater the risk, and vice-versa. Each record is assigned with a score that indicates how ``risky'' a record is. This score is determined by $\Pi_{i=k}^M(ATT-i)$, where $M$ is the user-specified maximum size of MSUs, and $ATT$ is the total number of attributes in the data set. The higher the score, the higher the risk. However, finding the MSUs is very challenging and SUDA is restricted to data sets with very small numbers of attributes. SUDA2 was proposed to deal with these drawbacks especially concerning the search space~\cite{manning2008recursive}. 

These three groups of measures are suitable for categorical attributes but not for continuous attributes as the number of uniques in a continuous attribute is usually large. Thus, several measures appropriate for this scenario are presented as follows, concerning the second main strategy for measuring identity disclosure risks in microdata. 

\subsubsection{Record linkage}

Population uniqueness presents limitations as a measure of identity disclosure risk. For example, it does not consider the characteristics of information possessed by intruders. Also, if a large number of sample and population uniques exists -- common when sets of QI have continuous attributes -- the number of uniques may not provide much information. Also, when the sampling fraction is small, it is difficult to estimate the number of population uniques accurately. Thus, measures of population uniqueness can be misleading. Record linkage aims to address these shortcomings. In the literature, record linkage is also known as identity matching~\cite{li2011identity} or fuzzy matching~\cite{nawaz2021fuzzy}.

Linking released records with target records can be employed either through direct matching using external data sets or indirect matching using existing data set~\cite{reiter2005estimating}. In both approaches, the data processor essentially mimics the behaviour of an intruder trying to match released records to target records. Since the data processor knows the real correspondence between original and de-identified records, it is possible to determine the percentage of correctly linked pairs. If the number of matched pairs is too high, the data set needs a robust de-identification before it can be released. 

The basic approach of record linkage is based on matching values of shared attributes. If common attributes share equal values in a pair of records and they are the only two records sharing such values, it is a matching pair. A non-matching pair happens when records differ in a common attribute value or multiple pairs of records exist sharing those same attribute values. Assuming two data sets, \textit{A} (original) with \textit{a} elements and \textit{B} (de-identified) with \textit{b} elements, the comparison space is the product of all possible record pairs $A \times B = \{(a, b): a \in A, b \in B\}$ which corresponds to the disjoint sets $M=\{(a, b): a = b, a \in A, b \in B\}$ and $U=\{(a, b): a \neq b, a \in A, b \in B\}$, where \textit{M} and \textit{U} correspond to the matched and non-matched sets respectively. Researchers studied the several types of record linkage for privacy risk assessment~\cite{domingo2004disclosure, hall2010privacy, domingo2016database, torra2017privacy}. Common strategies include probabilistic, distance and rank-based, outlined as follows.

\textbf{Probabilistic-based.} The goal of probabilistic-based record linkage~\cite{fellegi1969theory, jaro1989advances} is to assign a numerical value that reflects the similarity or dissimilarity of two records. Such a similarity is expressed as the ratio of two conditional probabilities that the pair of records have the same agreement pattern across the attribute of interest. Given a comparison vector $\gamma$ of the record pairs, the conditional probability that a pair is a match is $m(\gamma)=P(\gamma|(a, b) \in M) = P(\gamma|M)$ and a pair is a non-match corresponds to $u(\gamma)=P(\gamma|(a, b) \in U)=P(\gamma|U)$. Hence, a linkage rule defined as $R= m(\gamma) / u(\gamma)$, where $R \geq t_m$ a match is found and $R \leq t_u$ when a non-match is found, being $t_m$ and $t_u$ thresholds to be set.

\textbf{Distance-based.} Initially proposed by~\citet{pagliuca1999some}, distance-based record linkage aims to compute distances between records in original microdata and de-identified data. This requires appropriate distance metrics. Euclidean or Mahalanobis distance is used by~\citet{torra2006using}. Distance-based record linkage finds, for every protected record $b \in B$, an original record $a \in A$ which minimises the distance to $b$. Comparing distance and probabilistic-based, the former is more simple to implement~\cite{domingo2002distance}. Establishing the right distances for the attributes is the main difficulty in this approach. Furthermore, numerical attributes need to be normalised before the computation of the distance. 
    
\textbf{Rank-based.} Given the value of a de-identified attribute, the rank-based procedure validates whether the corresponding original value falls within an interval centred on the de-identified value, where the interval width is a rank~\cite{muralidhar2016rank, nin2008rethinking}. The main advantages of this approach are that no further scaling or standardisation is necessary. Also, while distance-based record linkage is commonly implemented using the minimum distance criterion, rank-based procedures use different criteria for selecting a match which is based on characteristics of the de-identification procedure.

The number of pairwise computations can be enormous even for small data sets. At the same time, the majority of similarity computations are dispensable given that most pairs are highly dissimilar and have no influence on the purpose. To avoid unnecessary computations a blocking phase is performed. In the blocking phase groups (blocks) of observations are formed using indexing or sorting. This technique selects a subset of record pairs from each block for subsequent similarity computation, ignoring the remaining pairs as highly dissimilar.~\citet{herzog2007data} present a set of record linkage case studies including a discussion on blocking.

\subsubsection{Outliers}

Extreme individuals are particularly easy to identify and therefore the disclosure risk is very high. Outlier detection is applicable for continuous QI where the objective is to identify at risk all the records for which the QI takes a value greater than a pre-defined quantile of the observed values. The outlier approach generates a fixed percentage of records at risk, depending on the subjective choice of the quantile.

Two approaches for outlier detection stems from the rank-based and standard deviation-based intervals presented by~\citet{mateo2004outlier}. Given the value of a de-identified attribute, it is verified if the corresponding original value falls within an interval centred on the de-identified value. The width of the interval is based on the rank of the attribute or its standard deviation. 
~\citet{truta2006global} also presented a standard deviation-based intervals approach for outliers detection. Besides rank-based and standard deviation-based approaches, the distance-based~\cite{vaidya2004privacy, templ2008robust} and density-based~\cite{foschi2011disclosure} are also used for the same purpose. Of all the above approaches, two of them are suitable for a multivariate setting~\cite{templ2008robust, foschi2011disclosure} while the remaining are for univariate setting.

\subsubsection{Clustering}

Measures based on clustering techniques are typically used when continuous QI are present in the microdata. Contrary to the common application of the clustering techniques which tries to find as many clusters as possible, the focus is to find clusters of size one, which indicates the higher risk. \citet{bacher2002re} showed how a standard hierarchical clustering algorithm can be applied to decide whether a record is safe or not. A limitation of standard algorithms is that they tend to find clusters with an equal variance which would only find records at risk on the tails of the distribution of the continuous attributes. Thus, the clustering algorithms may not be suitable for QI with skewed distributions~\cite{hundepool2012statistical}. A density-based approach appropriate to disclosure risk is proposed by~\citet{ichim2009disclosure}. If a certain record $r$ is very distant from nearest neighbours, it can be singled out. Contrarily, if nearest neighbours of $r$ are very close to $r$, an intruder cannot be certain about the match between $r$ and any of its neighbours. Such uncertainty increases with the density around $r$ which can be modelled by the density of records in a neighbourhood. An advantage of this local measure is its independence on the location of the records at risk, usually in the tails or centre of the distribution.

\subsection{Attribute Disclosure}

Attribute disclosure aims to assess an intruder's potential to correctly determine values of specific unknown attributes. The intruder may not precisely identify a particular data subject's record, but could infer a data subject sensitive values from released data, based on a set of sensitive values associated with the equivalence class of the individual.

In the case of more than one record representing the same individual in the data set, each equivalence class may not contain $k$ distinct individuals.~\citet{wang2006anonymizing} introduced the notion of $(X, Y)$-anonymity, where $X$ and $Y$ are disjoint sets of attributes. Suppose $x$ is a value on $X$; then, the de-identification of $x$ concerning $Y$ denoted as $a_Y(x)$ is the number of distinct values on $Y$ that occur with $x$. Such a measure illustrates how many values on $x$ are linked to at least $k$ distinct values on $Y$. If each value on $X$ corresponds to a group of individuals and $Y$ represents the sensitive attribute, each group is associated with a diverse set of sensitive values, making sensitive value inference difficult.

The $(\alpha, k$)-Anonymity~\cite{wong2006alpha} measure prevents attribute disclosure with the requirement that in any equivalence class, the frequency of a sensitive value is less than or equal to $\alpha$, where $0 < \alpha < 1$. For better protection, no single sensitive attribute can be dominant in an equivalence class. A very similar measure is proposed by~\citet{machanavajjhala2007diversity}, showing that intruders with more background knowledge can deduce sensitive information about individuals even without re-identifying them. $l$-diversity measure, indicates how many $l$-``well represented'' values are in each sensitive attribute for each equivalence class~\cite{machanavajjhala2007diversity}. Such definition corresponds to one of three types of ``well represented'': distinct $l$-diversity, which is identical to $p$-sensitive $k$-anonymity~\cite{truta2006privacy}. This distinct $l$-diversity is also applicable to identity disclosure when $k=l$, because each equivalence class contains at least $l$ records. The second type is the entropy $l$-diversity where the entropy of the distribution of the values of a sensitive attribute for each equivalence class is given by $log(l)$. Lastly, recursive $(c, l)$-diversity which indicates if the most frequent values do not appear too frequently and the least frequent values do not appear too rarely. An equivalence class is recursive $(c, l)$-diversity if $r_1<c(r_l + r_{l+1}+,... + r_m)$, where $m$ corresponds to the number of values of the sensitive attribute in an equivalence class and $r_i$ is the frequency of the $i^{th}$ most frequent value. When all the tuples of a given equivalence class have the same value for a sensitive attribute is called homogeneity of the values, which is when the attribute values are disclosed. 

A drawback of $l$-diversity is its applicability only to categorical sensitive attributes. In addition, it struggles with scenarios where data is skewed. Consider a patient data set where 95\% of individuals have Flu and 5\% of records have HIV; suppose that an equivalence class has 50\% of Flu and 50\% of HIV. Then, it is said that the equivalence class is distinct 2-diversity. Under those circumstances,~\citet{li2007t} introduced $t$-closeness that aims to measure how close the distribution of a sensitive attribute in an equivalence class is from the distribution of the attribute in the overall data set. In other words, $t$-closeness evaluates the distance between frequency distributions of sensitive attribute values, where $0 \leq t < 1$. The greater the distance, the greater the protection level. $\sigma$-Disclosure privacy~\cite{brickell2008cost} is very similar to $t$-closeness. It also focuses on the distances between the distributions of sensitive attribute values. However, the previous measure does not translate directly into a bound on the intruder's ability to learn sensitive attributes associated with a given QI. Therefore, $\sigma$-disclosure privacy uses a multiplicative definition. $\beta$-likeness~\cite{cao2012publishing} is related with the two previous measures, $t$-closeness and $\sigma$-disclosure privacy, however, it uses a relative difference measure. Suppose $p_{si}$ is the frequency of the sensitive value $s_i$ in the overall microdata and $q_{si}$ is the frequency of $s_i$ within any equivalence class. Then, $D(p_i, q_i)$ is defined as $(q_{si} - p_{si}) / p_{si}$ which describes the $\beta$.

In addition to identity disclosure, record linkage can also be used for attribute disclosure risk~\cite{SoriaComas2015AssessingDR}. An intruder that links the records in the de-identified data set to an external data set (that contains the background knowledge) attempts to assign an identity to the de-identified records. The linkage is usually based on a set of attributes that are common to both data sets. To maximise the accuracy of the linkage, the risk is assessed based on maximum-knowledge intruders. The objective is to perform a record linkage using only the $V-1$ attributes of the original data and de-identified data. 

An appropriate measure also for both identity and attribute disclosure types was proposed by~\citet{anjum2018efficient} called $(p, k)$-angelisation. This measure is applicable when the intention is to release two tables, a table with QI values and a second table with sensitive attributes values. The $p$ parameter indicates how many categories belong to certain records in every bucket and $k$ corresponds to the amount of $k$ tuples in each bucket. In other words, a single group containing $k$ records belongs to $p$ categories. The terminology of buckets is introduced later in Section~\ref{sec:PPT}.

\subsection{Summary}

Privacy risk measures quantify the degree of a privacy breach and thus, the amount of protection offered by privacy-preserving techniques. As such, privacy risk measures contribute to improving data subjects' privacy in the digital world. Below, we present the main measures from the literature regarding disclosure risk: identity and attribute disclosure. 

Identity disclosure risk is divided into two main groups according to QI data type: measures for categorical and numerical QI attributes. For categorical data, we should use measures based on uniqueness. Uniqueness aims to provide the number of records that possess unique combinations regarding selected QI. There are three common procedures: singling out, probabilistic modelling and special uniques. Singling out measures identify records that are typically easy to isolate by an intruder. Despite the high risk that these records possess, they could also be unique in the population data set. In such a case, unique records in the population data set means that they are the maximum risk. However, the population data set is not always known. Accordingly, the population frequencies can be estimated. To infer the population frequencies of a given QI combination from the corresponding sample frequency, a probabilistic model may be pursued. Lastly, an alternative strategy to defining disclosure risk is based on the concept of special uniqueness. The special unique measures must be used to determine with high certainty if a sample record is in fact unique, since a special unique usually present a higher probability of being a population unique than a sample unique which is not special. The determination of a special unique does not require the complete set of QI. It is impractical using uniqueness measures for numerical QI attributes since there will be many distinct combinations of QI. Under those circumstances, different measures should be applied. We recommended measures focused on outliers and clustering strategies.  

Besides the above measures suitable for identity disclosure, record linkage can be used to link records in de-identified data and records in original data to find how many are coincidentally providing the records at risk and need further protection. Additionally, this procedure can also be applied to indicate the possibility of increasing information about a data subject by adding new attributes to the data. Unfortunately, record linkage techniques may be difficult to apply and time-consuming for data processors when searching for a direct match, since it requires external data for that purpose. 



Table~\ref{tab:riskmeasures} summarises the privacy risk measures for each disclosure risk type with the respective main bibliographic references and the data type regarding QI. All these measures are applicable to the record-level. 

\begin{table}[ht!]
\centering
\scriptsize
\begin{tabular}{@{}ccccc@{}}
\toprule
\textbf{Disclosure risk}                     & \multicolumn{2}{c}{\textbf{Type}}                                       & \textbf{Privacy risk measures}     & \textbf{QI data type} \\ \midrule
\multirow{25}{*}{\textbf{Identity}}          & \multirow{16}{*}{Uniqueness} & \multirow{2}{*}{Singling out}            & $k$-Anonymity~\cite{samarati2001protecting}             & Categorical           \\ \cmidrule(l){4-5} 
                                             &                              &                                          & $k$-Map~\cite{el2008protecting}                            & Categorical           \\ 
                                            
                   \cmidrule(l){3-5} &                              & \multirow{10}{*}{Probabilistic modelling} & Poisson-gamma~\cite{bethlehem1990disclosure}                      & Categorical           \\ \cmidrule(l){4-5} 
                                             &                              &                                          & Poisson-log-normal~\cite{skinner1998estimating, skinner2002measure}                 & Categorical           \\ \cmidrule(l){4-5} 
                                             &                              &                                          & Ewens~\cite{ewens1990population}                              & Categorical           \\ \cmidrule(l){4-5} 
                                             &                              &                                          & Dirichlet-multinominal~\cite{takemura1999some}             & Categorical           \\ \cmidrule(l){4-5} 
                                             &                              &                                          & Logarithmic series~\cite{hoshino2001applying}                 & Categorical           \\ \cmidrule(l){4-5} 
                                             &                              &                                          & Pitman~\cite{hoshino2001applying}                             & Categorical           \\ \cmidrule(l){4-5} 
                                             &                              &                                          & Gaussian Copulas~\cite{rocher2019estimating}                   & Categorical           \\ \cmidrule(l){3-5} 
                                             &                              & \multirow{2}{*}{Special uniques}         & SUDA~\cite{elliot2002computational}                               & Categorical           \\ \cmidrule(l){4-5} 
                                             &                              &                                          & SUDA2~\cite{manning2008recursive}                              & Categorical           \\ \cmidrule(l){2-5} 
                                             & \multicolumn{2}{c}{\multirow{5}{*}{Outliers}}                           & Rank-based intervals~\cite{mateo2004outlier}               & Numerical             \\ \cmidrule(l){4-5} 
                                             & \multicolumn{2}{c}{}                                                    & Standard deviation-based intervals~\cite{mateo2004outlier, truta2006global} & Numerical             \\ \cmidrule(l){4-5} 
                                             & \multicolumn{2}{c}{}                                                    & Distance-based intervals~\cite{vaidya2004privacy, templ2008robust}           & Numerical             \\ \cmidrule(l){4-5} 
                                             & \multicolumn{2}{c}{}                                                    & Density-based intervals~\cite{foschi2011disclosure}            & Numerical             \\ \cmidrule(l){2-5} 
                                             & \multicolumn{2}{c}{\multirow{2}{*}{Clustering}}                         & Distance-based~\cite{bacher2002re}                     & Numerical             \\ \cmidrule(l){4-5} 
                                             & \multicolumn{2}{c}{}                                                    & Density-based~\cite{ichim2009disclosure}                      & Numerical             \\ \midrule
\multirow{9}{*}{\textbf{Identity/Attribute}} & \multicolumn{2}{c}{\multirow{4}{*}{Record linkage}}                     & Probabilistic-based~\cite{fellegi1969theory, jaro1989advances}                     & Both                  \\ \cmidrule(l){4-5} 
                                             & \multicolumn{2}{c}{}                                                    & Distance-based~\cite{pagliuca1999some, torra2006using}                & Both                  \\ \cmidrule(l){4-5} 
                                             & \multicolumn{2}{c}{}                                                    & Rank-based~\cite{muralidhar2016rank, nin2008rethinking}                         & Both                  \\ \cmidrule(l){2-5} 
                                             & \multicolumn{2}{c}{\multirow{3}{*}{$k$-Anonymity-based}}                &
                $(\alpha, k)$-Anonymity~\cite{wong2006alpha} & Categorical \\
                \cmidrule(l){4-5}
                & 
                \multicolumn{2}{c}{}              & $l$-Diversity~\cite{machanavajjhala2007diversity}                      & Categorical           \\ \cmidrule(l){4-5} 
                                             & \multicolumn{2}{c}{}                                & $(p, k)$-Angelisation~\cite{anjum2018efficient}              & Categorical           \\ \midrule
\multirow{5}{*}{\textbf{Attribute}}          & \multicolumn{2}{c}{\multirow{5}{*}{$k$-Anonymity-based}}                & $(X, Y)$-Anonymity~\cite{wang2006anonymizing}                 & Categorical           \\ \cmidrule(l){4-5} 
                                             & \multicolumn{2}{c}{}                                                    & $t$-Closeness~\cite{li2007t}                      & Categorical           \\ \cmidrule(l){4-5} 
                                             & \multicolumn{2}{c}{}                                                    & $\sigma$-Disclosure privacy~\cite{brickell2008cost}        & Categorical           \\ \cmidrule(l){4-5} 
                                             & \multicolumn{2}{c}{}                                                    & $\beta$-Likeness~\cite{cao2012publishing}                   & Categorical       \\ \bottomrule
\end{tabular}
\caption{Privacy risk measures at record-level for microdata with the main bibliographic references.}
\label{tab:riskmeasures}
\end{table}


\section{Privacy-Preserving Techniques} \label{sec:PPT}

The main challenge of de-identification is to discover how to release data that is useful for organisations, administrations and companies to make accurate decisions without disclosing sensitive information on specific data subjects. In other words, we are interested in the exploration of techniques that reduce the disclosure risk and still allows to perform statistical analysis and data mining tasks. Such a conflict between data privacy and utility has motivated research in the development of new privacy-preserving techniques or the refactoring of existing techniques. 

\begin{figure}[!ht]
   \centering
   \scriptsize
   \includegraphics[width=0.65\linewidth]{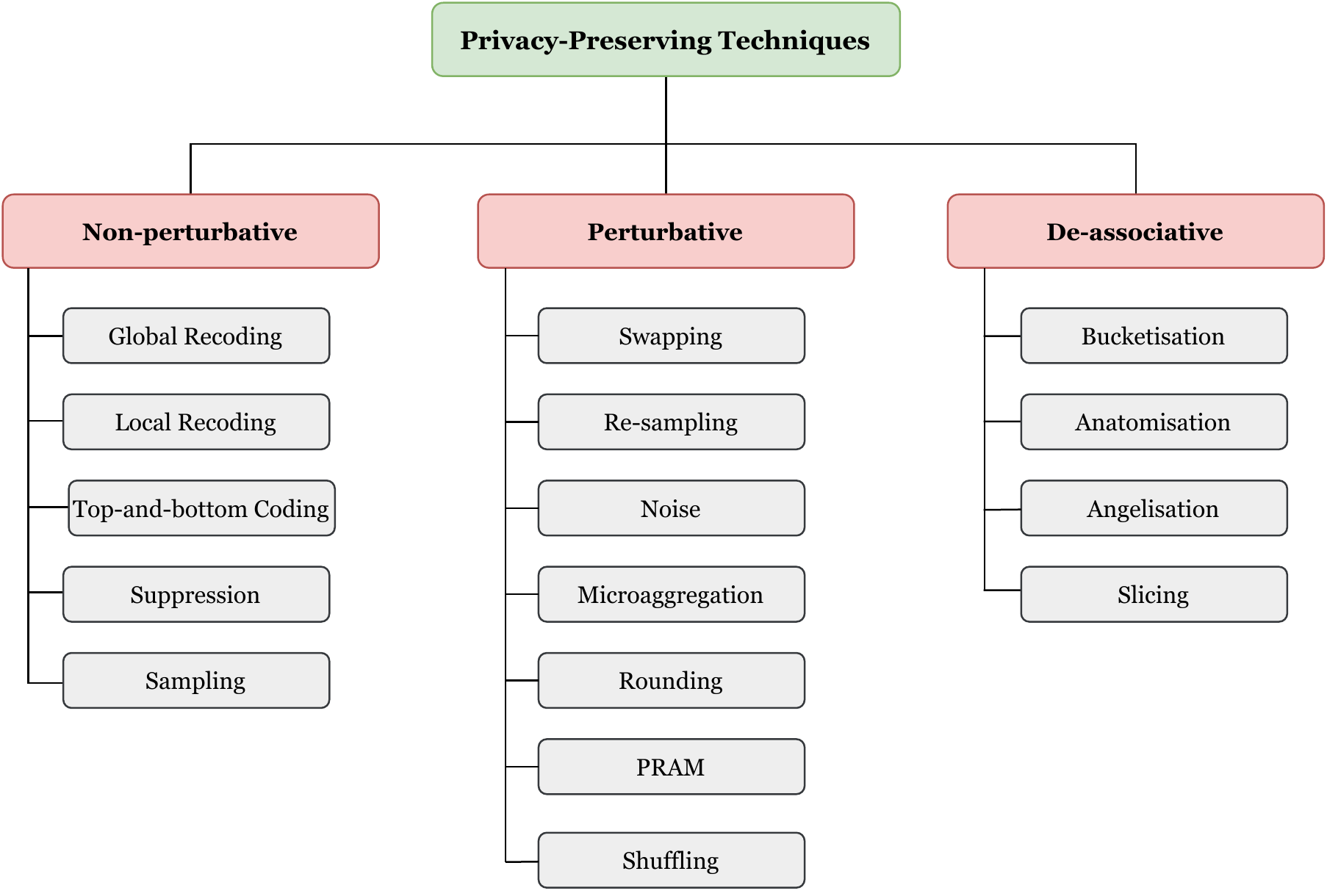}
 \caption{Taxonomy for privacy-preserving techniques in microdata.}
 \label{fig:tax}
\end{figure}

Willenborg and Waal~\cite{de1996view, willenborg2000elements} were among the first to present the principles of protecting microdata. They suggested a classification for privacy-preserving techniques according to the microdata characteristics. Their proposed taxonomy includes both non-perturbative and perturbative techniques. The former involves the reduction of detail or even suppression of information; the latter relates to distortion of information. When considering the background of an intruder, the distinction between non-perturbative and perturbative techniques is important. For instance, the inconsistencies provoked by perturbative techniques can generate special interest in an intruder to identify the records that may have been changed and to recover the original values. With non-perturbative techniques, this is not possible since inconsistencies are not generated. 
Despite the high use and discussion in the literature on the previous group of techniques, there are others that we assume to be part of SDC. We call this new category de-associative techniques. The main goal of de-associative techniques is to break the relationship between QI and sensitive attributes, releasing two separate tables instead of just one with both QI and sensitive attributes. Figure~\ref{fig:tax} provides a general overview of the main privacy-preserving techniques, which are discussed below. 

\subsection{Non-perturbative} The objective of this category is to reduce the amount of information in data by reducing the level of detail or partially suppressing information in original data preserving the truthfulness. Thus, non-perturbative techniques do not modify the data. We describe global recoding, local recoding, top-and-bottom coding, suppression and sampling techniques. 

\subsubsection{Global Recoding}

This technique is often known as a generalisation or full-domain generalisation. The essence of recoding is to combine several categories to create new and more general categories. Considering a microdata $T$, the application of global recoding in a categorical attribute $V_i$, i.e., replacing with a more generalised $g$ category, results in a new $V_i'$ with $|D(V_i')| < |D(V_i)|$, where $D$ is the domain of $V_i$. For a continuous attribute, $V_i$ is replaced by a discretised version o $V_i$. By intuition, the smaller the sizes of intervals in $g$, less information loss.
The main objective of global recoding is to divide the tuples of $T$ into a set $E$ of disjoint equivalence classes and then transform the QI values of the tuples in each equivalence class to the same format. Applying global recoding means that the QI values of all equivalence classes obey the principle that there cannot be two equivalence classes with overlapping tuples. Each equivalence class is also known as QI-group. This technique is used heavily in literature and by the statisticians~\cite{willenborg1996statistical, willenborg2000elements, hundepool2012statistical}.

\subsubsection{Local Recoding}

While global recoding uniformly recodes the values across the microdata set, local recoding recodes into broader intervals or categories when necessary~\cite{takemura1999local}. The replacement can be partial, only some occurrences of $V_i$ are replaced with $g$. Local recoding can potentially reduce the distortion in the data, by replacing $V_i$ only in a neighbourhood of the data. In general, the difference between global and local recoding is that with global recoding, all values of the attribute have the same domain level. On the other hand, with local recoding, the values are generalised to different domain levels. As global recoding affects the entire data set, the risk can be reduced, however, it can have a greater cost in terms of information loss. With local recoding, this problem can be contained~\cite{xu2006utility}.

\subsubsection{Top-and-bottom Coding}

Top-and-bottom coding is a special case of recoding. This technique is applied to continuous or categorical ordinal attributes. A top recoding covers values of an attribute above a specified upper threshold $\theta_u$ where frequencies tend to become smaller toward the top of the attribute range. Similarly, a bottom coding covers values below another threshold $\theta_l$. However, determining the appropriate threshold is not a simple task~\cite{willenborg2000elements}. 

\subsubsection{Suppression}

This technique suppresses data from microdata so that they are not released, or replaces values with a missing value (\textit{NaN}) or a special character ('*' or '?'). Suppression has been used since the 80s~\cite{cox1980suppression} and is common in combination with global recoding~\cite{willenborg1996statistical, hurkens1998models, samarati2001protecting}. There are three common levels of suppression, outlined below. 
\begin{itemize}[nosep]
    \item[--] \textbf{Cell suppression.} 
    This approach is also known as local suppression. The goal is to replace one or more values of an attribute $V_i$ in a record $R$ with a missing value or a special character. In a tuple of QI $t$, if $t$ does not occur frequently enough, one or more values in $t$ can be suppressed.
    In contrast to global recoding that is applied to the entire microdata, cell suppression is only applied to a particular value in a particular record. This means that cell suppression does not change the attribute definitions, because it does not affect the coding of any attribute. However, cell suppression significant reduces the predictive performance~\cite{ohno2002effects}. 
    
    \item[--] \textbf{Tuple suppression.}
    It is also known as record suppression. The main idea is to hide the whole tuple $t$ from the released data, which may contain both non-sensitive and sensitive information. Nevertheless, if some records are removed from the data, it may disturb its truthfulness.

    \item[--] \textbf{Attribute suppression.}
    An attribute $V_i$ is no longer released for the whole data set. This approach could be very useful when a categorical attribute has many distinct values which typically affects the predictive performance in data mining applications~\cite{carvalho_compromise2021} or when two attributes are highly correlated.
    
\end{itemize}

\subsubsection{Sampling}

Also known as subsampling, it is a common technique to protect census microdata where the original set corresponds to the entire population~\cite{skinner1994disclosure}. Instead of releasing the original microdata set, a sample $S$ set is published. Sampling is suitable for categorical attributes, however, for continuous attributes may be inappropriate owing to the fact of many distinct values being present as this technique does not perturb continuous attributes~\cite{hundepool2012statistical}. If a continuous attribute $V_i$ is present in a public data set, unique matches are very probable. For that reason, the publishing of a sample of microdata set with continuous attributes should be accomplished with other privacy-preserving techniques.


\subsection{Perturbative} Perturbative techniques distort data before release. These must be used so that the statistics computed on the perturbed microdata set do not differ significantly from statistics obtained on the original data set. Swapping, re-sampling, noise, microaggregation, rounding, PRAM and shuffling are examples of perturbative techniques. 

\subsubsection{Swapping} 

\citet{dalenius1982data} propose a data swapping technique, in which the main idea is to exchange the values of certain attributes across records. There are two types of swapping, record or data swapping and rank swapping. Even though data swapping was originally proposed for categorical attributes, its rank swapping variant is also applicable to continuous attributes. The notion of this technique is to swap pairs of records that are similar on a set of attributes but belong to different sub-domains.  

\begin{itemize}[nosep]
    \item[--] \textbf{Data swapping.} 
        In the application of this technique, the microdata is considered a matrix. Suppose a set of \textit{n} individuals, each one containing the values of \textit{m} attributes, which are quasi-identifiers and sensitive attributes. The microdata is then represented as $n \times m$ matrix, corresponding to the original matrix $m_o$. Data swapping will maps such matrix into another matrix, $m_e$. The resulting matrix can be released to micro-statistics or be used to produce macro-statistics. The $m_e$ is \textit{t-order equivalent} with $m_o$ to preserve the \textit{t-order} statistics. A \textit{t-order equivalence} is a frequency count table in which $t$ attributes are involved. A \textit{1-order equivalence} involves one attribute. A set resulted by swapping contains \textit{n} individuals, \textit{t} attributes, \textit{r} individuals in each equivalence class and \textit{k} of \textit{r} individuals have their values swapped. 
        
    \item[--] \textbf{Rank swapping.} 
    Was originally introduced by~\citet{moore1996controlled} for ordinal attributes, but~\citet{domingo2001disclosure} show that this technique can also be used for numerical attributes. The difference from the original procedure is the restriction of the range for which each value can be swapped. This strategy aims to limit the distortion. The methodology of this approach is as follows: values of attribute $t_i$ are ranked in ascending order; then each ranked value of $t_i$ is swapped with another ranked value randomly chosen within a restricted range. The range typically corresponds to the amount of swapped values. For instance, the rank of two swapped values cannot differ much from \textit{p\%} of the total number of records, where the $p$ is commonly between 0 and 20.~\citet{nin2008rethinking} propose a new record linkage suitable for rank swapping and two new variants called \textit{rank swapping p-distributions} and \textit{rank swapping p-buckets} which are effective to the new record linkage method.  
\end{itemize}

Data swapping has many advantages, namely \textit{i)} it removes the relationship between the record and the individual; \textit{ii)} it can be used in one or more sensitive attributes without disturbing the non-sensitive attributes; \textit{iii)} no non-sensitive attributes are deleted; \textit{iv)} provides protection to the rare and unique values, and \textit{v)} it is not limited to the type of attributes. This technique has also some drawbacks, for instance, arbitrary swapping can produce a large number of records with unusual combinations. If the swapping is not random, the function to determine the records and the attributes to be swapped requires significant time and computer resources. Also, this technique can severely distort the statistics on any sub-domain, for example, mean and variance for the income of nurses. Lastly, data and rank swapping do not prevent attribute disclosure, since they only reorder data. For example, when an intruder knows which record in the original data set has the highest income, the intruder will simply have to look for the highest income in the swapped data set to get the exact value~\cite{hundepool2012statistical}. Swapping has been applied for Japanese population census microdata as a potential technique to replace the deletion of unique records~\cite{ito2014data, Ito2018}. Although swapping has desirable properties, it is not suitable for attributes with few distinct values, since a value can be swapped with the same or similar values.  

\subsubsection{Re-sampling}

Re-sampling was first applied for tabular protection~\cite{heer1993bootstrap}, but this technique can also be applied to microdata sets~\cite{hundepool2012statistical}. The re-sampling strategy uses the bootstrap method with a replacement that consists in repeatedly taking small samples and calculating the average of each sample. Formally, re-sampling takes $t$ independent samples $S_1$, ..., $S_t$ of size $n$ of the values of an original attribute $V_i$. Each sample must be sorted with the same ranking criterion. Build the masked attribute by taking as first value the average of the $S_1$, as second value the average of the $S_2$, and so on. 



\subsubsection{Noise}

Protecting personal data with noise can be done by additive or multiplicative noise. Additive noise has been studied and used extensively since the 80s~\cite{spruill1983confidentiality,kim1986method, tendick1991optimal, brand2002microdata}. Noise is also known as randomisation. Four main procedures have been developed for additive noise~\cite{hundepool2012statistical}:

\begin{itemize}
    \item[--] \textbf{Uncorrelated noise addition.}
    Adding noise means that the vector of observations $x_j$ for the $j^{th}$ attribute in the original microdata is replaced by the vector $z_j$, where $z_j=x_j+ \epsilon_j$ and $\epsilon_j$ denotes normally distributed errors derived from $\epsilon_j \sim N(0,\sigma_{\epsilon_j}^2)$ with $Cov(\epsilon_T,\epsilon_l)=0$ for all $t \neq l$ (white noise). The higher the $\sigma$ value, the greater the range of the generated values.
    
    \item[--] \textbf{Correlated noise addition.}
    Adding correlated noise is based on the generation of an error matrix $\epsilon ^*$ under the restriction $\sum^*=\alpha \sum$ (correlated noise), which implies $\sum _Z=\sum+\alpha\sum=(1+\alpha)\sum$. All elements of the covariance matrix of the perturbed data diverge from those of the original data by a factor $1+\alpha$. The difference between correlated and uncorrelated is is that the covariance matrix of the errors in this approach is proportional to the covariance matrix of the original data. Using correlated noise addition produces a data set with higher analytical validity than uncorrelated noise addition.
    
    \item[--] \textbf{Noise addition and linear transformation.}
    Ensuring additional linear transformations mean that the sample covariance matrix of the changed attributes is an unbiased estimator for the covariance matrix of the original attributes. This concept uses a simple additive noise on the $p$ original attributes $Z_j=X_j+\epsilon_j$, for $j=1, ..., p$ with covariances of the errors proportional to those of the original attributes. The shortcoming of this approach is that it cannot be applied to discrete attributes as do not preserve the univariate distributions of the original data.

    \item[--] \textbf{Noise addition and non-linear transformation.}
    To bypass the shortcoming of previous approach, the non-linear transformations was proposed in combination of additive noise~\cite{sullivan1989use}. The application of such an approach is very time-consuming and requires expert knowledge on the microdata set and the algorithms~\cite{hundepool2012statistical}.
    
\end{itemize}

In some cases, it is preferable to work with multiplicative noise. For example, when additive noise has constant variance. This phenomenon causes small values to be strongly perturbed and large values weakly perturbed. Multiplicative noise was then proposed to circumvent this drawback~\cite{nayak2011statistical}. Suppose a matrix $X$ of the original numerical data and $W$ the matrix of continuous perturbation attributes with expectation 1 and variance $\sigma_w^2>0$. The resulting perturbed data $X^a$ is obtained by $X^a=W\odot X$, where $\odot$ corresponds to Hadamard product, an element-wise of matrix multiplication.

In addition to $k$-anonymity, differential privacy is in the spotlight of research. Both $k$-anonymity and differential privacy are the most discussed subject in the literature in the data privacy domain. However, there are some substantial differences between the two. Although $k$-anonymity is used to guide privacy-preserving techniques, we notice that it is a measure that gives the number of individuals that share the same information and usually is used in data transformed with generalisation and suppression. In the case of differential privacy, it is a method used to usually apply Laplace noise addition. Besides that, differential privacy does not require any assumptions on the intruder's background knowledge. 
The initial formulation of differential privacy was proposed by~\citet{dwork2006differential} for an interactive setting. A randomised query function satisfies differential privacy if the result of an individual's information is the same whether or not that individual's information is included in the input for the analysis. Notwithstanding, researchers proposed extensions of differential privacy for the non-interactive setting, namely for microdata sets that could be used for any analysis~\cite{soria2014enhancing}. We should clarify that, in the scope of this survey, differential privacy methods are grouped as noise-based methods.

\subsubsection{Microaggregation}

This technique was initially proposed to be used for continuous attributes~\cite{domingo2002security, domingo2002practical}, but microaggregation has been extended to categorical data~\cite{torra2004microaggregation}. 
The $n$ records in microdata set $T$ are partitioned in $g$ groups of $k$ or more individuals. Groups are formed using a criterion of maximal similarity, for instance, Euclidean distance can be used. The value $v_i$ of a continuous attribute $V$ in record $r$ is replaced by an aggregate value, usually the average, mode or median of the group to which $r$ belongs. Some variants exist for microaggregation besides this approach for continuous attributes: \textit{i)} fixed group size~\cite{defays1993panels, domingo2002security} which requires that all groups be of size $k$; \textit{ii)} variable group size~\cite{domingo2002security, domingo2002practical, laszlo2005minimum} that allows groups to be of size $\geq k$; \textit{iii)} univariate~\cite{defays1993panels} deal with multi-attribute microdata sets by microaggregating one attribute at a time, \textit{iv)} multivariate~\cite{domingo2002security} deal with several attributes at a time, \textit{v)} categorical attributes~\cite{torra2004microaggregation, martinez2012semantic}, \textit{vi)} optimal~\cite{hansen2003polynomial, laszlo2009approximation} aims to find a grouping where groups have maximal homogeneity and size at least $k$, and \textit{vii)} heuristic~\cite{defays1993panels, domingo2006efficient} to deal with multivariate microaggregation. 

Microaggregation involves three main criteria: how the homogeneity of groups is defined, the clustering algorithms used to find the homogeneous groups, and the determination of aggregated function. Typically, this technique works better when the values of the attributes in the groups are more homogeneous. For that reason, the information loss caused by the replacement of values with common values will be smaller than in cases where groups are less homogeneous.

\subsubsection{Rounding}

Rounding has been used for a very long time as a technique in this context~\cite{dalenius1981simple}. Its objective is to replace the original values of attributes with rounded values. For a given attribute $X_i$, rounded values are chosen among a set of rounding points defining a rounding set. The set of rounding points ${p1, ..., p_r}$ can be determined by the multiples of a base value $b$: $p_i=b \times i$ for $i=1,... ,r$. Thus, an original value $x_i$ of $X$ is replaced with the rounding point. In multi-attribute microdata, rounding is typically performed one attribute at a time (univariate rounding). Multivariate rounding is also possible~\cite{willenborg2000elements}. The operating principle of this technique makes it suitable for continuous data.

\subsubsection{Post RAndomisation Method (PRAM)}

PRAM is a technique used for categorical data and was proposed by \citet{gouweleeuw1998post}. The values of one or more categorical attributes are re-coded with a certain probability and such re-coding is done independently for each of the records. As PRAM uses a probability mechanism, an intruder cannot be sure whether certain matches corresponds to the correct individual.
Suppose $\xi$ is a categorical attribute in the original set to which PRAM will be applied and $X$ is denote the same attribute in the perturbed set. Both $\xi$ and $X$ has $K$ categories (or factor levels). The probabilities are defined by $p_{kl} = P(X = l|\xi = k)$, which mean that the probability that an original score $\xi = k$ is re-coded with the score $X = l$. Such probabilities are called \textit{transition probabilities}. In the case of PRAM for a single attribute, the transition matrix has a size of $k \times k$, which results in a Markov matrix.

The transition probabilities must be chosen appropriately since can occur in certain situations unlikely combinations like a 7-year-old girl being identified as being pregnant. In similar cases, the transition matrix can be designed in such a way that those transitions are not possible (transition probability set equal to 0). Despite this disadvantage, PRAM is especially useful when a microdata set contains several attributes and applying other de-identification methods, such as global recoding, local suppression and top-and-bottom coding, would lead to too much information loss~\cite{hundepool2012statistical}.  



\subsubsection{Shuffling}

Data shuffling was proposed by Muralidhar and Sarathy~\cite{muralidhar2003theoretical, muralidhar2003a} and it is a variation of swapping~\cite{swapping_variations2004}. The difference between swapping and shuffling is that this technique replaces sensitive attributes by generating new data with similar distributional properties. Suppose that \textit{X} represents sensitive attributes and \textit{S} non-sensitive attributes. It is generated new data \textit{Y} to replace \textit{X} using the conditional distribution of \textit{X} given \textit{S}, $f(X|S)$. The generated values are also ranked, as in rank swapping. Each \textit{X} value is replaced with another \textit{X} value with the rank that corresponds to the rank of the \textit{Y} value. When compared to the swapping technique, data shuffling guarantees a higher level of utility and a lower level of disclosure risk~\cite{muralidhar2006swap}. Similarly to data swapping and rank swapping, shuffled data has a potential risk of attribute disclosure~\cite{hundepool2012statistical}. Furthermore, data shuffling requires a ranking of the whole data set, which can be computationally costly for large data sets with millions of records~\cite{muralidhar2006data}.






\subsection{De-associative} The principal objective of this category is to create buckets in order to break the correlation between QI and sensitive attributes. The basis of de-associative techniques is to create QI-groups with at least $k$ records. In the de-associative category are included bucketisation, anatomisation, angelisation and slicing techniques.

\subsubsection{Bucketisation} 

The term bucketisation derives from the created buckets (often known as partitions) when the values are recoded into less specific values in such a way that each record in the bucketised set corresponds to multiple sensitive values. Therefore, an intruder cannot distinguish the tuples in the same bucket. The bucketised data consist of a set of buckets with permuted sensitive values. This technique publishes the QI values in their original format, thus it is not difficult for an intruder to find out whether an individual is present in the published data or not. For that reason, bucketisation does not prevent membership disclosure. 

To prevent both identity and sensitive attribute disclosure, and to preserve the data utility, some researchers propose certain modifications to the bucketisation approach, namely not releasing the bucketised data with the QI in the original form~\cite{li2017cross, li2020local}. The QI can be generalised and then bucketised. Thus, the size of buckets is reduced as well as information loss. The tuples are partitioned into equivalence groups that satisfy the desired protection, then divides the generalised tuples into buckets to break the connection between QI values and sensitive values. However, we believe that this approach is only suitable for scenarios where the data controller knows the complete background knowledge of the adversary, otherwise, there are still unique QI combinations that can lead to identity disclosure. 
 
\citet{zhang2007} proposed a permutation-based approach to reduce the association between QI and sensitive attributes suitable for microdata and query-based databases. This technique aims to partition the tuples into several groups so that each group has at least $l$ different sensitive attribute values. Such an approach is identical to the definition of bucketisation and for that reason, we assume henceforth that permutation corresponds to bucketisation.

\subsubsection{Anatomisation}

\citet{xiao2006anatomy} proposed a new technique called Anatomy that follows the bucketisation principles but, instead of permuting sensitive values, aims to publish two separate tables: a QI table (QIT) and a sensitive table (ST). Consider a microdata $T$ which contains a set of $V$ quasi-identifiers either categorical or numerical and a categorical sensitive attribute $S$. Given an equivalence class with $m$ \textit{QI-groups}, this technique produces a QIT in the form of ($V_1, V_2$, ..., $V_i$, \textit{Group-ID}) and a ST in the following format (\textit{Group-ID}, $S$, $Count$). Each \textit{QI-group} involves at least $l$ tuples. 
For each \textit{QI-group} and each distinct $S$ value $v$ in $QI_j$ ($1 \leq j \leq m$), the ST has a record of the form: ($j, v, c_j(v)$), where $c_j (v)$ is the number of tuples $t \in QI_j$ with $t[d + 1] = v$ ($d$ is the number of QI). This approach releases the QI and sensitive attributes in two separate sets conserving the unique common attribute, the \textit{Group-ID}. Hence, when an intruder tries to join the two tables, he will not be able to associate the sensitive value to the right individual. 

Although anatomisation increased its popularity in data privacy, this technique is vulnerable to background knowledge attacks and it is only be applied to limited applications. For that reason, it should be used in combination with other privacy-preserving techniques. For instance, anatomisation can be combined with bucketisation, where the values in QIT are permuted as well the values in the ST~\cite{He2012Permutation}. Besides that, it focuses on the de-identification of microdata with a single sensitive attribute. To solve this drawback, a multiple sensitive bucketisation~\cite{zhiwei2017research} was proposed to partition the microdata into QIT and ST and to make sure each sensitive attribute has at least $l$ diverse values. This approach follows the division into two tables like anatomisation, and thus, we consider that their proposal is a combination of anatomisation with bucketisation.~\citet{ye2017anonymization} also discuss the problem of secure releasing data when sensitive data contains multiple attributes and also combining anatomisation with bucketisation. The difference from the previous approach is that the QI values in the QIT are randomly permuted. 


\subsubsection{Angelisation}

As with anatomisation, angelisation aims to release two separate tables.~\citet{tao2009angel} propose a very similar approach called Angel and, such an approach starts by dividing the microdata into batches $B_1, B_2,$ ..., $B_b$ where each batch is a set of tuples in $T$ and the sensitive attribute distribution in each batch $B_i$ satisfies a certain objective de-identification principal $P$. Then, creates another partition but into buckets, $C_1, C_2,$ ..., $C_e$ where each bucket is a set of tuples in $T$ and contains at least $k$ tuples, in which $k$ is a controlling parameter of the degree of protection. The angelisation corresponds to the publication of any pair of bucket and batch partitions. Therefore, given a batch and bucket partitions of the $T$, an angelisation of $T$ corresponds to a pair of a batch table (BT) $\langle$\textit{Batch-ID, S, Count}$\rangle$ and a generalised table (GT) that has all QI attributes with the column of \textit{Batch-ID} where all tuples in the same bucket $C_i$ have equivalent generalised QI values. 
While anatomisation releases a QIT and ST, angelisation releases a BT and GT. Moreover, anatomisation releases the QI directly, which is a concern when disclosing the precise QI values. However, if there is a specific scenario where it is needed the release of the original values, then angelisation can be employed instead of anatomisation, as this approach also allows the direct publication of QI values. 


\subsubsection{Slicing}

Due to the break of correlation between QI and sensitive attributes caused by previous techniques,~\citet{li2010slicing} propose the slicing approach which groups several QI with the sensitive attribute, preserving attributes correlations. The intuition of slicing is the partition of a microdata set vertically and horizontally. Vertical partitioning groups attributes into columns based on the correlation of the attributes, i.e., each column contains a subset of highly correlated attributes. Horizontal partitioning groups tuples into buckets. For each bucket, values in each column are randomly permuted to remove the linking between different columns. Formally, in a microdata $T$, an attribute partition corresponds to several subsets of attributes. A tuple partition consists of several subsets of $T$ and each subset of tuples is a bucket. A slicing of $T$ is given by an attribute partition, a column generalisation and a tuple partition.


As slicing was initially proposed for single sensitive attribute~\citet{han2013sloms} propose to vertically partition multiple attributes into several ST and on QIT. In each table, the tuples are partitioned into equivalence classes and the QI values of each equivalence class are recoding satisfying a desired level $k$. The sensitive values of each ST are sliced and bucketised to achieve $l$ diverse values. Besides that, this technique can be used in combination with other techniques, for example with anatomisation for multiple sensitive attributes~\cite{susan2016anatomisation}. 
A different version, based on overlapped slicing, aims to duplicate a sensitive attribute and put it into the QI column~\cite{budiardjo2019privacy}. This approach will increase data utility since there is more attribute correlation. Furthermore, it works for multiple sensitive attributes.

\subsection{Summary}

Many privacy-preserving techniques were proposed to limit disclosure of confidential information. To clarify the main concept of each technique, we presented three groups to help data controllers in decision making. Table~\ref{tab:ppt_summary} summarises the main principles of each privacy-preserving technique and highlights which data type techniques are intended for. Generally, non-perturbative aims to reduce information detail without distortion. Perturbative methods aim to distort information by creating uncertainty. Lastly, de-associative techniques aim to break the correlation of QI and sensitive attributes. Table~\ref{tab:ppt} presents the main advantages and disadvantages for these groups of privacy-preserving techniques. 

\begin{table}[ht!]
\centering
\scriptsize
\begin{tabular}{@{}clll@{}}
\toprule
\multicolumn{2}{c}{\textbf{Privacy-preserving techniques}}         & \multicolumn{1}{c}{\textbf{Data type}} & \multicolumn{1}{c}{\textbf{Principle}} \\ \midrule
\multirow{5}{*}{\textbf{Non-perturbative}} & Global recoding       & Both                                       &   Combines several categories to form more general categories.                                     \\
                                           & Local recoding        &  Both           & Recodes into broader categories only some values.                                       \\
                                           & Top-and-bottom coding &   Both                                     &  Replaces values above or bellow of a defined threshold.                                      \\
                                           & Suppression           &  Both          &   Deletes cells/rows/attributes or replace them with special characters.                                     \\
                                           & Sampling      &    Both        &  Selects a sample of the original microdata.                                      \\ \midrule
\multirow{7}{*}{\textbf{Perturbative}}     & Swapping      &   Both                                   &   Exchanges the values of certain attributes across records.                                    \\
                                           & Re-sampling           &   Both         &  Takes independent samples using bootstrap and averages the samples.                                     \\
                                           & Noise        &  Numerical                & Replaces the values by adding/subtracting/multiplicating random values.                                       \\
                                           & Microaggregation      &  Both      &    Groups similar values and assigns an aggregated value to the group.                                    \\
                                           & Rounding           & Numerical          & Replaces values with rounded ones determined by the multiples of a base.                                       \\
                                           & PRAM                  &                  Categorical       &  Reclassifies the values according to the Markov matrix.                                      \\
                                           & Shuffling      & Numerical              &   Uses a regression model to determine which values are exchanged.                                     \\ \midrule
\multirow{4}{*}{\textbf{De-associative}}   & Bucketisation         &           Categorical      &  Bucketises data and permutes the sensitive values.   \\
                                           & Anatomisation         &   Categorical     &   Bucketises data and publishes a QIT and a ST.                                      \\
                                           & Angelisation  &    Categorical   &   Divides data into batches and then buckets, and publishes a BT and a GT.                                     \\
                                           & Slicing     &    Categorical      &   Partitionates data vertically and horizontally.                                     \\ \bottomrule
\end{tabular}
\caption{Summary of each privacy-preserving technique including data type and main principles.}
\label{tab:ppt_summary}
\end{table}

\begin{table}[ht!]
\centering
\scriptsize

\begin{tabular}{@{}cll@{}}
\toprule
{\textbf{Privacy-preserving techniques}} & \multicolumn{1}{c}{\textbf{Advantages}}      & \multicolumn{1}{c}{\textbf{Disadvantages}}                                                                                                                     \\ \midrule
\textbf{Non-perturbative}                                                             & \begin{tabular}[c]{@{}l@{}}- Does not disturb data structure;\\ \\ - Unique combinations may disappear. \end{tabular}                                                                                                                               &   \begin{tabular}[c]{@{}l@{}} - Reduces the detail of information;\\ \\ - High generalisation level and many suppressed values \\ destructs data utility.\end{tabular} \\ \midrule

\textbf{Perturbative}                                                                 & \begin{tabular}[c]{@{}l@{}}- Creates uncertainty around the values;\\ \\ - New combinations may appear. \end{tabular}                                                                                                                                                   & \begin{tabular}[c]{@{}l@{}} - It may creates inconsistencies;\\ \\ - Extreme values require great quantity of distortion.\end{tabular}                                                            \\ \midrule
\textbf{De-associative}                                                               & \begin{tabular}[c]{@{}l@{}}- Publishes QIs in their original form;\\ \\ - Breaks the relationship between QIs and \\ sensitive attributes.\end{tabular} 

& \begin{tabular}[c]{@{}l@{}}- Presents high disclosure risk when QIs are the original ones;\\ \\ - The swapped sensitive values could interfere with its true \\ meaning and patterns leading to inaccurate results.\end{tabular}                                                                      \\  \bottomrule
\end{tabular}

\caption{Main advantages and disadvantages of each privacy-preserving technique.}
\label{tab:ppt}
\end{table}

As mentioned, privacy-preserving techniques must be applied according to data characteristics and based on the evaluation of privacy risk and data utility. So far, we reviewed privacy risk measures and techniques for data transformation. In the following section, we present data utility measures for different end-use cases.

\section{Data utility}~\label{sec:utility}

De-identified data must be protected in such a way that it can be used for diverse purposes. However, taking into account all data uses is impracticable. Therefore, we must consider data interpretability when data is used for general purposes by using information loss measures and predictive performance measures for data mining/machine learning tasks. Thus, we can classify data utility measures into two main groups: information loss and predictive performance.

Greater the perturbation applied to microdata via privacy-preserving techniques, greater the distance between the original and the de-identified data set; consequently, greater information loss.  
The main idea for loss information measures is to compare records between the original and de-identified data and compare statistics computed from both data sets~\cite{domingo2001disclosure, hundepool2012statistical, fletcher2015measuring}. Measures of information loss allow the data processor to assess how much harm is being inflicted to the data by a particular privacy-preserving technique. In other words, such measures allow evaluating whether a data set is still analytically valid/comparable after de-identification.



Many measures were proposed in the literature regarding information loss of the de-identified data for general purposes. 
The information loss measures can be divided into three groups: distance/distribution comparisons, penalty of transformations through generalisation and suppression, and statistical differences.
An example for the former case is the discernibility measure~\cite{bayardo2005data} which sums up the squares of equivalence class sizes. KL-Divergence~\cite{kullback1951information} is also well known, and measures differences in the distributions of equivalence class sizes over the same attribute. It indicates how much information is lost after changing the probability distributions in the data. In addition to these measures, covariance comparisons can be used for the same purpose~\cite{domingo2001disclosure}. For the second case, the number of generalisation steps can be given by the average size of the equivalence classes~\cite{machanavajjhala2007diversity, nergiz2007thoughts}. In the other hand, the minimal distortion~\cite{samarati2001protecting} charges a penalty to each generalised or suppressed value. A similar measure, GenILoss~\cite{iyengar2002transforming}, penalises a specific attribute when it is generalised. 
The last case is based on the frequencies of the equivalence classes. For example, comparing the number of missing values in the data sets indicates the degree of information loss usually caused by specific privacy-preserving techniques. Information loss also can be quantified through the changes in statistics. For instance, by comparing means, variances and correlations in the data sets.


Regardless of the previous general information loss measures, the data processor may know the end-use of de-identified data. Accordingly, when data is intended to be used for data mining tasks, several researchers have proposed to evaluate the utility of the de-identified data in terms of data mining workloads~\cite{lefevre2006workload, brickell2008cost, fletcher2015measuring, carvalho_compromise2021}. Since a de-identified data set should support further analysis, predictive models are built from the de-identified data set, and then the prediction accuracy of the models is used to represent the usefulness of the de-identified data set. However, it is unclear what type of data mining tasks will be performed on the de-identified data. In data mining tasks, utility references predictive performance -- the closer the evaluation results obtained between the original and the de-identified data, the more utility is preserved. Classification is a common task studied for this purpose that aims to predict nominal attributes. Typical measures used to assess predictive performance include Precision and Recall~\cite{Kent_prec_acc}, Accuracy, F-score~\cite{Rijsbergen_fmeasure}, Geometric Mean~\cite{kubat1998machine} and AUC (Area Under the ROC Curve)~\cite{weng2008new}. Besides these measures,~\citet{iyengar2002transforming} proposes a Classification Metric that measures the classification error on the training data by penalising transformations done by suppression or generalisation in which the record's class is not the majority class. Nonetheless, when the task involves predicting a numeric value, regression measures are used to evaluate the predictive performance. Common measures for this scenario include Mean Squared Error, Root Mean Squared Error and Mean Absolute Error. 

Both classification and regression tasks are supervised learning approaches in which utility/performance can be measured by the power of discriminating class labels. But unsupervised learning, namely clustering, is also a common approach to evaluate the quality of the de-identified data and no class labels are available~\cite{fletcher2015measuring, fung2008framework}. Intuitively, clustering refers to grouping records in such a way that similar records are grouped together and dissimilar records are grouped in different clusters. Common measures for clustering analysis includes, for instance, Rand index~\cite{rand1971objective}, Davies-Bouldin index~\cite{davies1979cluster}, Fowlkes-Mallows index~\cite{fowlkes1983method} and Silhouette~\cite{rousseeuw1987silhouettes}. Table~\ref{tab:utility} summarises the main bibliographic references for the two main groups of data utility measures.

\begin{table}[ht!]
\centering
\scriptsize
\begin{tabular}{@{}ccc@{}}
\toprule
\multicolumn{2}{c}{\textbf{Data utility measures}}                                   & \textbf{Main references}         \\ \midrule
\multicolumn{2}{c}{\multirow{3}{*}{\textbf{Information loss}}}                       &  \citet{bayardo2005data}, \citet{kullback1951information}, \\
\multicolumn{2}{c}{}                                                                 & \citet{machanavajjhala2007diversity}, \citet{samarati2001protecting},                          \\
\multicolumn{2}{c}{}                                                                 & \citet{iyengar2002transforming}, \citet{nergiz2007thoughts}                        \\ \midrule

\multirow{4}{*}{\textbf{Predictive performance}} & \multirow{2}{*}{\textbf{Supervised}}   & \citet{Kent_prec_acc}, \citet{Rijsbergen_fmeasure}, \citet{kubat1998machine},                       \\
                                            &                                        &  \citet{weng2008new}, \citet{iyengar2002transforming}                             \\ \cmidrule(l){2-3} 
                                            & \multirow{2}{*}{\textbf{Unsupervised}} & \citet{rand1971objective}, \citet{davies1979cluster},                             \\
                                            &                                        & \citet{fowlkes1983method}, \citet{rousseeuw1987silhouettes}                              \\ \bottomrule
\end{tabular}
\caption{Data utility measures and main bibliographic references.}
\label{tab:utility}
\end{table}

As an objective, for maximum data utility, one should strive for de-identified data to be as similar to the original data as possible. However, guaranteeing maximum utility will result in a lower data protection level, which as previously referred, could affect both data subjects and organisations. Thus, ensuring an optimal level of data privacy and utility requires great effort in the application of privacy-preserving techniques. In the following section, we present existing studies on the effectiveness of such techniques concerning data privacy and utility in terms of predictive performance.  


\section{Studies on the effectiveness of privacy-preserving techniques} \label{sec:studies}

The application of privacy-preserving techniques is not always trivial. Although a certain technique seems to improve privacy, for example, to reduce the granularity of data, may in fact keep the level of privacy for some occurrences intact due to the extreme values. Moreover, reducing the detail of information may negatively impact data utility. For this reason, it is extremely important to conduct studies in order to evaluate the effectiveness of such techniques.  

In this section, we discuss existing studies on the impact of privacy-preserving techniques for both data privacy and utility regarding predictive performance. Also, we cover available software, their principles and implementation details.

\subsection{Impact on data privacy}

Privacy-preserving techniques have been proposed to limit the disclosure of private information. Improvements and new proposals popped after the suggestion of $k$-anonymity measure. Such a measure was initially used to measure identity disclosure in data that were transformed with generalisation and suppression techniques~\cite{samarati2001protecting}. However, it was proven that both techniques are not sufficient to protect disclosure of sensitive data~\cite{machanavajjhala2007diversity, li2007t}. A suggestion to increase the individuals' privacy is to suppress values that have high disclosure risk~\cite{orooji2019improving}. 

The introduction of noise is one of the most used technique in the perturbative group. A well-known conclusion is the challenge of generating a perturbed data set that remains statistically close to the original data. Typically, the more close the perturbed data is to the original, the less confidential that data set becomes. On the opposite side, the more distant the perturbed data set is from the original, the more secure it is. However, the utility of the data set might be lost when the statistical characteristics of the original data set are lost~\cite{mivule2013utilizing}.   

Although the distortion applied by differential privacy has been widely considered to be a robust standard, recently,~\citet{muralidhar2020epsilon} presented an empirical study that compares two approaches of differential privacy via microaggregation. Their experimental results show that a fixed $\epsilon$ does not guarantee a certain level of confidentiality. Thus, this method is not good for microdata releases, which challenges the previous theoretical guarantees.


Microaggregation is also very used for perturbation of microdata sets and has been enhanced in terms of disclosure risk. For instance,~\citet{fadel2021microaggregation} presented very recently a heuristic approach to apply microaggregation that aims to reduce the disclosure risk when compared with other approaches. 
On the other hand, existing works in microaggregation shows that this technique either produce a low degree of with-in cluster homogeneity or fail to reduce the amount of noise independent of the size of a data set and for these reasons,~\citet{iftikhar2019publishing} propose an interesting approach that uses microaggregation for generating differentially private data sets. 

De-associative techniques have been also explored and improved to protect individuals' privacy. The principal drawback of these types of techniques is the publishing of the QI tables in their raw form. But also the disclosure risks for some absolute facts which would help the intruder to ﬁnd invalid records in the transformed data set resulting in the disclosure of confidential information. Therefore,~\citet{hasan2016effective} proposed the combination of slicing and data swapping to decrease the attribute disclosure risk. By swapping the values, the published data contains no invalid information such that the intruder cannot disclose the individual privacy. On the other hand,~\citet{sari2020asenva} states that breaking the two sets of attributes produces more records than the original data set. The authors' proposal use generalisation and suppression in the QIT, then sensitive values are aggregated on a ST and QI attributes are summarised. Thus, the number of records in the sensitive table is reduced. However, conclusions are limited as privacy risk and utility were not determined. But, theoretically, this approach provides better protection.

The main challenge for an organisation is to apply the optimal privacy-preserving techniques that reduce disclosure risks with minimal information loss. Currently, there are no ``one size fits all'' approaches to data privacy. Nevertheless, some guidelines should be used to limit disclosure risk when releasing data to the public, industries or researchers. Data release must follow a defence strategy that uses both technical and non-technical approaches to data privacy~\cite{murray2021privacy, humbert2019survey}. 

A non-technical approach aims to understand the intended use of data since it would be naive to assume that all malicious attacks against individuals' privacy have been discovered at this stage. Also, an intruder may possess external information to join with a released data set to re-identify individuals in the de-identified data set. To address such issue, non-technical approaches should be considered to provide risk-limiting solutions. As such, data sets should not be made freely available without barriers. An example of a barrier is the identification of the study that data receivers will conduct by specifying the data they require and how data will be used to achieve their goals. Thus, the cost to a potential disclosure risk by an intruder will be raised in terms of effort. Furthermore, it allows the data controller to filter out data receivers without malicious intent, but whose work intends to violate the data subjects' privacy. In addition, another non-technical defence is an agreement in which data receivers commit not to de-identify any data.

Regarding technical approaches, one of the main issues is the understanding of which privacy-preserving techniques are most appropriate for certain situations. Many of these techniques require trial and error to decide which parameters to configure and to find the acceptable trade-off. Therefore, it is important to decide which privacy-preserving technique is most appropriate to apply to a given data set for a specific application.

\subsection{Impact on predictive performance}

The application of privacy-preserving techniques are many times employed through de-identification algorithms, for example, Incognito~\cite{lefevre2005incognito}, Mondrian~\cite{lefevre2006mondrian}, 
and many other de-identification algorithms that are surveyed by~\citet{fung2010privacy}. The aim of such algorithms is to apply techniques such as global recoding and local suppression with minimal information loss. In general, information loss is measured during the generalisation process. We notice that the studies in predictive performance rely often upon data sets produced with these de-identification algorithms. 

In the context of supervised learning tasks, namely classification, several studies are aiming to prove the efficiency of generalisation and suppression in predictive analysis. A common approach is to transform the original data using such techniques in QI attributes that are highly identifiable to satisfy privacy constraints and using $k$-anonymity to evaluate the privacy of transformed data. Such transformation originates several de-identified data sets and they all are evaluated according to the models' performance. Many examples~\cite{iyengar2002transforming, wang2004bottom, fung2005top, inan2009using,buratovic2012effects} show higher classification errors for data sets produced by optimising the loss metric and reach the maximum possible value at higher values of $k$. In general, they prove that de-identification level increase leads to proportional degradation of predictive performance. However, it is still possible to protect individuals' privacy while maintaining predictive performance with both techniques.

A different approach was introduced by~\citet{brickell2008cost} which focus on semantic definitions to quantify the attribute disclosure. Besides $k$-anonymity, the authors also used $l$-diversity, $t$-closeness and proposed a new measure to capture the adversarial knowledge gain. Whereas the previous works used a defined set of QI, this study uses different sets of QI. 
In most cases, trivial de-identification, i.e., removing all QI or all sensitive attributes, provides equivalent performance and better privacy guarantees than common generalisation and suppression. Such results challenges previous works and their conclusions. However, we believe that this result depends on the selected set o QI, for instance, if a higher number of attributes are suppressed, the de-identified data set intuitively will not have much utility. 

\citet{lefevre2006workload} presented a suite of de-identifications methods to generate de-identified data based on target workloads, which consists of several data mining tasks, including both classification and regression tasks. In contrast to the previous proposals which uses de-identification algorithms to avoid identity disclosure, this experience focus also on attribute disclosure. 
In general, the derivations of Mondrian algorithm outperforms the~\citet{fung2005top} de-identification algorithm. 
However, the previous methods suppress too many values~\cite{li2011information}. Therefore, the level of generalisation can be determined by the distribution of the attributes and then use cell suppression to remove locally detailed information. Such a method was proven to be more accurate in classification than~\citet{lefevre2006workload} proposal. 

Besides generalisation and suppression, a few studies also include data sets with noise
~\cite{vanichayavisalsakul2018evaluation, carvalho_compromise2021}. Contrary to~\citet{vanichayavisalsakul2018evaluation}, the conclusions of~\citet{carvalho_compromise2021} point towards a noticeable impact of such techniques in predictive performance, especially with noise. However, noise is the one that presents a low re-identification risk level. 
Beyond that, the former uses several de-identification algorithms, while the latter tested different parameters in privacy-preserving techniques without a de-identification algorithm. 

The single application of noise was also performed by some researchers~\cite{mivule2012towards, mivule2013comparative, zorarpaci2020privacy}. The experimental studies allow us to conclude that the level of noise does affect the classification error. Such results are expected, as the higher $\epsilon$, the less is the noise~\cite{lee2011much}; therefore, the private data is more closely to the original data.
Although the noise adds uncertainty to the intruder in the re-identification ability, it may result in additional data mining bias~\cite{wilson2003protecting}. The introduced bias could severely impact the ability of knowledge discovery. It is usually related to the change in variance, the relationships between attributes or the underlying attribute distribution.~\citet{wilson2003protecting} shows that additive noise has a lower impact on classification compared to other types of noise. 

Microaggregation was also tested in terms of predictive performance. A comparative study of several microaggregation approaches w.r.t predictive analysis and $k$-anonymity measure shows that the prediction accuracy of a classifier based on a de-identified data set is not always worse than baseline~\cite{lin2010comparison}. For instance, some results show higher accuracy when compared to the baseline due to the reduction of variance in the de-identified data set. 

Besides the regression workload previously presented~\cite{lefevre2006workload},~\citet{ohno2002effects} also presents a study focused on regression tasks using both suppression and generalisation. 
However, the predictive performance is evaluated concerning the number of suppressed cells. With the minimum privacy guarantees, the results show a difference compared with the original data. But, a slightly higher privacy level results in approximately the same predictive performance. 
More recently,~\citet{liu2019uhrp} presented a new approach to reduce the uncertainty introduced by the privacy-preserving techniques. The idea is to represent features using vectorising functions in a de-identified instance or original instance. Experiments show the regression model trained with de-identified data can be expected to do as well as with original data set under certain features representations.

Regarding unsupervised tasks, clustering methods have been used to evaluate the privacy-preserving techniques. Some approaches convert the problem into classification analysis, wherein class labels encode the cluster structure in the data and then evaluate the cluster quality on the de-identified data. For instance, Fung et al.~\cite{fung2008framework, fung2009privacy} define the de-identification problem for cluster analysis using generalisation. After transforming the data, the clusters in the original data set should be equal to those in the de-identified data set. In general, the cluster quality degrades as the de-identification threshold increases. However, the results suggest that is possible to achieve a reasonable level of de-identification without compromising cluster quality.

\citet{oliveira2010privacy} proposed a new distortion technique to numerical attributes in order to meet the desired privacy level in clustering analysis. The experiments include additive noise, multiplicative noise and rotation noise that is defined based on an angle $\theta$. Their technique shows a misclassification between 0\% and 0.2\%. In particular, multiplicative noise achieved the best values for accuracy and privacy level in most experiments. In general, the experiments shows that is possible to achieve a good compromise between privacy and accuracy. 



\subsection{Available Software}

Many privacy tools have been developed to help data controllers or any other user in decision making for any purpose. Beyond the transformation of microdata through privacy-preserving techniques, some tools allow different configurations assessments of such techniques, enabling the evaluation of the achieved privacy and utility level.

$\mu$-ARGUS~\cite{de1996argus} was the pioneering tool in privacy-preserving publishing. It is a free open source software that provides a user interface and is currently in version 5.1.3~\cite{muargus}. This tool allows the application of techniques such as global recoding, top-and-bottom coding, local suppression, PRAM, noise addition and microaggregation. The individual risk estimation is based on sampling weight~\cite{benedetti1998individual} and $k$-anonymity. 
Furthermore, $\mu$-ARGUS allows the production of de-identified data sets suitable for different purposes, namely, for scientific and public use files.

UTD Anonymisation ToolBox~\cite{utd} is available for download and further implements all three groups of privacy-preserving techniques. Cornell Anonymisation Toolkit~\cite{cornell} also provides an interface and is free for download. Cornell tool uses generalisation to transform the data. Both tools depicts the utility and re-identification risk. Moreover, these tools are research prototypes and have scalability issues when handling large data sets. 


Also, a free open source software for scientific and public use files is the R package called sdcMicro~\cite{sdcmicro}. This package supports several privacy-preserving techniques, both non-perturbative and perturbative. For risk estimation, sdcMicro implements for example the SUDA2, $k$-anonymity, log-linear models, among others. Besides that, this package provides the utility between the de-identified data and the original data and quantifies the information loss. The package also features a user-friendly interface~\cite{kowarik1sdcmicrogui} that allows non-experts in the de-identification process to gain insight into the impact of various privacy-preserving techniques and reduces the burden that that software can be used.

ARX Data Anonymisation Tool~\cite{prasser2020flexible}, often known as ARX is one of the most used tools for data privacy. This software is free and has a simple and intuitive interface~\cite{arx}, which supports wizards for creating transformation rules and visualisations of re-identification risks. A wide range of privacy-preserving techniques is supported in ARX as well as many measures for both re-identification risk and data utility. Regarding the data utility, ARX further optimises output data towards suitability as a training set for building learning models. ARX is also available as a library with an API that provides data de-identification for any Java program. 

Another free open source privacy tool is Amnesia~\cite{amnesia}, which provides software and an online dashboard. Amnesia allows the selection of generalisation level and uses $k$-anonymity for disclosure risk. This tool produces many possible solutions, and it shows the distribution of values and provides statistics about the data quality in the de-identified data set. Moreover, Amnesia allows transforming relational and transactional databases into de-identified data.

All the previous tools require the attribute terminology in advance, which is a static approach. A dynamic tool called Aircloack~\cite{aircloack} has emerged to help data controllers by giving them access to all the underlying data, and dynamically adapting the de-identification to the specific query and data requested. The resulting answer set is fully de-identified. Both non-perturbative and perturbative techniques are implemented Aircloack. This tool has a free version for students based on querying system and a full version for organisations.

Although some of the presented tools are more intuitive and user-friendly, they require prior knowledge of the SDC process. Also, for example, sdcMicro requires knowledge of the R programming language. These tools are useful for data protection but they are not trivial. Table~\ref{tab:tools} summarises the discussed privacy tools and their characteristics. 



\begin{table}[ht!]
\centering
\scriptsize
\begin{tabular}{@{}lccccccc@{}}
\toprule
\textbf{Privacy tool} & \textbf{Open source} & \textbf{Web app} & \textbf{Non-perturbative} & \textbf{Perturbative} & \textbf{De-associative} & \textbf{Privacy assessment} & \textbf{Utility assessment} \\ \midrule
\textbf{$\mu$-ARGUS}  &     \checkmark  &    $x$  &    \checkmark   &  \checkmark  &  $x$   &  \checkmark   &    $x$                   \\
\textbf{UTD}         &     \checkmark       &    $x$     &       \checkmark      &         \checkmark    &   \checkmark   &    \checkmark   &   \checkmark \\
\textbf{Cornell}  &     \checkmark     &    $x$     &   \checkmark  &    $x$     &    $x$   &    \checkmark     &    \checkmark     \\
\textbf{sdcMicro}     &     \checkmark     &   $x$     &   \checkmark    &  \checkmark   &    $x$    &    \checkmark     &  \checkmark   \\
\textbf{ARX}     &    \checkmark    &   $x$  &    \checkmark&     \checkmark      &    $x$    &    \checkmark      &    \checkmark      \\
\textbf{Amnesia}      &   \checkmark      &  \checkmark       & \checkmark       &  $x$    &     $x$  &    \checkmark       &     \checkmark    \\
\textbf{Aircloack}    &   $x$      &    \checkmark     &   \checkmark    &          \checkmark      &   $x$     &    \checkmark       &   \checkmark    \\ \bottomrule
\end{tabular}
\caption{Available privacy tools along with their main characteristics.}
\label{tab:tools}
\end{table}

\section{Open Issues} \label{sec:openissues}

Regardless of de-identification procedures currently implemented in the data privacy context, the expertise required to successfully perform these operations is high. With growing interest and importance of data privacy, and the exponential growth of available data, we expect this area of research and application to grow even further. As such, a linchpin to the widespread use of these methods concerns the automation of these procedural pipelines. For example, to our knowledge, there is still no solution that automates the application and optimisation of privacy-preserving techniques to achieve proper levels of privacy and data utility. These decisions may not be an easy task especially for a user without background knowledge in data protection, which could lead to dangerous decisions. To address such an issue, a potential solution to explore is the introduction of Automated Machine Learning (AutoML) solutions~\cite{Hutter2018} in this context, with the aim of simplifying the de-identification process for data mining tasks, and allowing the widespread application of robust procedures for de-identification and data sharing.

In addition, several open issues are urgent and require attention in upcoming years. In this section, we provide an overview of some problems and challenges that frequently coexist with the privacy-preserving data publishing research area. We briefly introduce problems such as dealing with data that is modified over time and republished, also the plethora of data that is currently easy to obtain, and finally dealing with privacy when the data is in several parties.

\subsection{Dynamic data}

One of the biggest challenges for the wider dissemination of privacy-preserving strategies is the management of dynamic data. This type of data is dynamically changed over time. If a set of privacy-preserving techniques are applied to dynamic data, the privacy-preserving problem would not be successfully solved. Particularly, the confidential information would not be properly protected while some records are added, deleted or modified. Typically, the information loss increases over time. Most of the existing strategies for centralised publication focus on static microdata. Specifically, they are restricted to only one-time publication and do not support republication. It is a future challenge the development of flexible, interactive and adaptive privacy-preserving techniques. 

A couple of proposals have been emerged to limit the disclosure risk in re-publication~\cite{xiao2007m, wang2010anonymizing}. However, location privacy should be enhanced for current and future privacy of releases such as in 5G and social networks~\cite{liao2018location}. Besides the location itself, the location information also includes when and for how long it was visited. Therefore, it is crucial the protection of spatio-temporal information. To avoid background knowledge and homogeneity attacks, the time attribute may be added in the de-identification of the location attribute~\cite{liu2020dynamic}. In continuous data release, outliers are also a problem. Outliers detection techniques should continuously monitor the cross-correlation between the releases. 


\subsection{Big data}

The dramatic increase of digital data, being gathered and shared currently, often cause the growing number of attributes causing the exponential increase of the domain. Under those circumstances, we face the so-called ``the curse of dimensionality'' which may cause information loss in de-identified data sets. Some approaches were suggested to mitigate the curse of dimensionality~\cite{lefevre2006mondrian, kifer2006injecting}. 
Additionally, big data variety may add further constraints to the de-identification procedure since techniques such as global recoding are highly dependant on data variations~\cite{zigomitros2020survey}. Thus, dealing with complex data is a challenge. 

The massive volume of confidential data being harvested by data controllers makes it essential to use the cloud services not just to store the data, but also to process them on the cloud premises.~\citet{domingo2019privacy} presents a survey that covers technologies for privacy-aware outsourcing of storage and processing of confidential data to clouds which includes the traditional privacy-preserving techniques and cryptographic methods. 
The main research challenge pointed by the authors is when constructing big data by merging data sets with overlapping subjects. 
Another limitation is the application of cryptographic methods in cloud computing which is at an early stage. Although it ensures high data protection, their application reduces utility and add computation overhead. 

\subsection{Distributed data}

Nowadays, due to the need for massive amounts of real-time data, centralised solutions are moving towards more decentralised ones. When data is in different parties, the approach to deal with this issue is often through collaborative data publishing in which multiple data providers share their data for general purposes or data mining tasks. In such a case, the de-identification data is given by each provider.

This topic has been explored, but the rapidly growing volume of data forces the exploration of new security measures. For instance, in a distributed scenario, where each organisation owns a set of raw individual data, de-identification operations are applied equally in all parties, which requires one party to act as leader, to synchronise the de-identification process~\cite{mohammed2010centralized}. However, it is important to include additional security measures. Currently, there are new proposals towards a solution based on multiple collaboration of different parties along with a cryptographic mechanism~\cite{lyu2020towards}. One of the main concerns in distributed systems is the balance between disclosure cost and computation cost.


\section{Conclusion}\label{sec:conslusion}

Privacy concerns are extremely important when confidential information disclosure is considered. As such, it is fundamental to protect data before making it available for any purpose. Many definitions of data privacy have been suggested. However, privacy can be defined as the prevention of unwanted information disclosure. In this article, we present a formulation of the de-identification problem for microdata. We exhaustively present the privacy risks to take into account before data is shared or released. We also extensively present the state of the art of the privacy-preserving techniques by proposing a new taxonomy for the existing approaches grouping them into \textit{i)} non-perturbative, \textit{ii)} perturbative, and \textit{iii)} de-associative. Nevertheless, the application of such techniques may provoke the destruction of data utility. As such, we explore existing measures of utility for general purposes and data mining tasks. Furthermore, we analyse several studies showing the impact of privacy-preserving techniques for both data privacy and utility. However, in this survey, we specifically present studies of the effectiveness of these techniques in predictive performance. 

Since the 2000s data privacy has been in the research spotlight either in suggesting new privacy measures and techniques or in evaluating them. The introduction of laws and regulations in data protection and the ability in private information disclosure have reinforced the need for new suggested solutions. Many privacy-preserving techniques were proposed, each one for specific purposes and data characteristics. But, we highlight the high importance in combine these techniques to guarantee high levels of data protection. Such a level is measured through the risks of private information disclosure and several measures were suggested for this purpose. However, diverse scenarios should be examined as it is very difficult to predict who the intruder is or the information he/she possesses.
Notwithstanding, high protection typically leads to lower data utility. Consequently, it is essential to experimentally evaluate the impact of de-identification by building predictive models from the de-identified data and observing how it performs on testing cases. Few works~\cite{fung2005top, lefevre2006workload, fung2008framework} have actually conducted such experiments, although many~\cite{iyengar2002transforming, bayardo2005data, brickell2008cost} addressed the classification task. We stress the relevance in reproducing experimental studies in regression tasks and for clustering analysis as the end use of data is often unknown.

Although we do not address strategies for others types of data, we strongly recommend some articles, for instance the survey of~\citet{zigomitros2020survey} on relational data. For trajectory microdata,~\citet{fiore2019privacy} present the solutions proposed for protecting this type of data form intruder attacks.   
A very recent survey on heterogeneous data was introduced by~\citet{cunha2021survey}. The authors propose a privacy taxonomy that establishes a relation between different types of data and suitable privacy-preserving strategies for the characteristics of those data types. Despite that,~\citet{wagner2018technical} surveys privacy measurements for several privacy domains where privacy-enhancing technologies (PETs), including privacy-preserving techniques, can be applied. 

A final remark on the certainty of this being a very relevant field for many years to come, especially considering the growing worries about data privacy and the need for ensuring such characteristics for increasing data needs for development and applications in machine learning or more generally, artificial intelligence.


\begin{acks}

The work of Tânia Carvalho is supported by Project “POCI-01-0247-FEDER-041435 (Safe Cities)” and financed by the COMPETE 2020, under the PORTUGAL 2020 Partnership Agreement, and through the European Development Fund (EDF). The work of Nuno Moniz is financed by National Funds through the Portuguese funding agency, FCT - Fundação para a Ciência e a Tecnologia, within project UIDB/50014/2020. The work of Luís Antunes is supported by EU H2020-SU-ICT-03-2018 Project No. 830929 CyberSec4Europe (cybersec4europe.eu). Pedro Faria is supported by Project "DataInvalue - NORTE-01-0247-FEDER-069497".

\end{acks}

\bibliographystyle{ACM-Reference-Format}
\bibliography{sample-authordraft}



\end{document}